# AFM manipulation of gold nanowires to build electrical circuits

M. Moreno-Moreno,[1,†] P. Ares,[1,†] C. Moreno,[2] F. Zamora,[2,3] C. Gómez-Navarro,[1,3] J. Gómez-Herrero*[,1,3]

[1]Departamento de Física de la Materia Condensada, Universidad Autónoma de Madrid, Madrid E-28049, Spain.

[2]Departamento de Química Inorgánica and Institute for Advanced Research in Chemical Sciences (IAdChem), Universidad Autónoma de Madrid, Madrid E-28049, Spain.

[3]Condensed Matter Physics Center (IFIMAC). Universidad Autónoma de Madrid, Madrid E-28049, Spain.

[†]These authors contributed equally to this work.

* E-mail: julio.gomez@uam.es

**ABSTRACT:** We introduce Scanning-Probe-Assisted Nanowire Circuitry (SPANC) as a new method to fabricate electrodes for the characterization of electrical transport properties at the nanoscale. SPANC uses an atomic force microscope manipulating nanowires to create complex and highly conductive nanostructures (paths) that work as nanoelectrodes allowing connectivity and electrical characterization of other nanoobjects. The paths are formed by the spontaneous cold welding of gold nanowires upon mechanical contact leading to an excellent contact resistance of ~9 $\Omega$/junction. SPANC is an easy to use and cost-effective technique that fabricates clean nanodevices. Hence, this new method can complement and/or be an alternative to other well-established methods to fabricate nanocircuits such as Electron Beam Lithography (EBL). The circuits made by SPANC are easily reconfigurable and their fabrication does not require the use of polymers and chemicals. In this work, we present a few examples that illustrate the capabilities of this method, allowing robust device fabrication and electrical characterization of several nanoobjects with sizes down to ~10 nm, well below the current smallest size able to be contacted in a device using the standard available technology (~30 nm). Importantly, we also provide the first experimental determination of the sheet resistance of thin antimonene flakes.

**INTRODUCTION**

Innovation in micro- and nano-electrical circuitry has countless implications in both fundamental research and daily life. Current challenges in this field are the design and assembly of devices incorporating emerging materials and/or architectures and the understanding of their properties. Electrical characterization of nanomaterials frequently implies the fabrication of metal electrodes. The standard option of choice to accomplish this task is Electron Beam Lithography (EBL),[1-3] a Scanning Electron Microscopy (SEM) based technology developed in the sixties.[4] EBL comprises the following steps: **i)** Deposition of a polymer mask, usually poly(methyl methacrylate) onto the sample surface. **ii)** Drawing of the custom circuit onto the polymer using a computer-controlled focused electron beam that cracks the polymer chains. **iii)** Removal of the marked region/s using an organic solvent. **iv)** Deposition of a metal layer. **v)** Final removal of the polymer (lift-off) leaving the custom circuit on the surface. EBL is a well-established technique with thousands of users worldwide. Commercial EBL systems rarely connect objects with sizes below 30 nm, which is good enough for multiple applications in nanotechnology.[5] However, it presents some drawbacks, as the need of an expensive computer-controlled SEM, the exposition of the samples to vacuum and chemical agents that might damage the nanoobjects, and the presence of residues left by the polymer mask. In addition, the circuits made by EBL are not reconfigurable: once they have been fabricated it requires a tremendous effort, if possible at all, to modify or upgrade them.

Atomic Force Microscopy (AFM), developed in the eighties,[6] is a newer technique than EBL. The "heart" of AFM is a very sharp tip at the end of a microcantilever that acts as a force sensor. Unlike SEM, AFM can work in very different environments: liquids, air ambient atmosphere, ultrahigh vacuum at cryogenic temperatures, etc. Thanks to its great versatility and affordable price compared to other nanoscale imaging techniques such as electron microscopy setups, AFM has become a common tool used routinely in many laboratories all over the world. Apart from imaging with resolution down to the atomic range,[7] AFM has also nanolithography modes. In the nineties, Dip-Pen lithography or local oxidation[8-10] were proposed for the fabrication of nanocircuits. More recently, other AFM-based fabrication procedures taking advantage of thermomechanical[11] and field emission[12] effects have been proposed. However, these techniques are restricted to very few special applications and do not deliver nanocircuits with the low electrical resistances provided by EBL.[13]

Carbon nanotubes based circuitry was developed as an attempt to characterize the electrical transport properties of nanoobjects.[14-16] However, the dependence of the nanotubes electronic gap on the chirality, the variation of their electrical resistance upon mechanical deformation and, more importantly, the high electrical resistance observed in nanotube junctions (inherent to systems with strong covalent bonding as carbon nanotubes) hampered the use of this technique. In 2010 Yang Lu *et al.* predicted that cold welding of metallic nanowires will have potential applications for electrical connectivity at the nanoscale.[17]

In our work, we introduce a new technique based on AFM that allows fabricating highly conductive and complex nanocircuits by manipulating and cold welding gold nanowires with the AFM tip. We have named this technique Scanning-Probe-Assisted Nanowire Circuitry (SPANC). Our first results demonstrate several basic features of SPANC: low electrical contact resistance between NW-NW and NW-nanoobject, high reproducibility and stability of the so-assembled nanocircuits and device fabrication. Then, we present examples illustrating the capabilities of this new AFM-based nanoelectrode fabrication technique: the robustness of the created circuits down to low temperature, the reconfigurable character of the electrodes, and the possibility of fabricating several-electrode stand-alone devices, which can be inserted in conventional experimental setups for further electrical characterization. In addition, we describe two experiments showing the ability to fabricate devices with electrical contacts for nanoobjects with sizes down to ~10 nm and the potential of SPANC for molecular electronics. We also consider fabrication of circuits on other insulating substrates and in a wide range of humidities. Finally, we briefly discuss possible applications of SPANC that in the near future will broaden the catalogue of existing techniques for nanofabrication.

**RESULTS**

Fig. 1a describes the basic steps required for the realization of nanocircuits by SPANC: **i)** Fabrication of the sample and micrometer-sized electrodes. **ii)** Deposition of gold nanowires (Au NWs) from a water suspension by drop casting (see details in Methods). **iii)** AFM manipulation of the Au NWs to form continuous paths. To accomplish this task, we first image the sample in amplitude modulation mode (AM-AFM), which ensures low tip-sample forces avoiding unwanted NW motion. Then, we bring the tip into hard contact with the substrate and we move it along a predefined trajectory manipulating the Au NW. Finally, we lift the tip back to AM-AFM mode to image the results of the manipulation. Additional details can be found in Methods and Supporting Information (SI1). When two NWs are brought into mechanical contact, as a consequence of the high surface-area-to-volume ratio, they cold weld spontaneously.[17] Despite the diameter of our NWs was about ten times larger than those used in [17], the cold welding between wires was very simple. However, we found that the optimum geometry for cold welding was head-to-side. On the contrary, welding wires in the side-to-side configuration was always more difficult. We also measured the lateral force required to move/weld two wires, finding that is in the order of 100 nN (see Fig. S1-3). According to Lu *et al.* this force should help to remove the CTAB surfactant facilitating the welding process.

In what follows, we describe three relevant circuit topologies which are commonly used in nanocircuitry. We can classify them according to the number of Au NW electrodes employed to contact the nanoobject (see Fig. 1b): **1) One** Au NW electrode. In this case, one NW path connects one microelectrode with the nanoobject, the second electrode is a mobile conductive AFM tip that closes the circuit. **2) Two** Au NW electrodes. In this configuration, two paths are fabricated to make the electrical contact between two microelectrodes and the nanoobject.

This setup is usually discarded for nanoobjects with low electrical resistance, as the electrical contact resistance of the circuit elements can be much higher than the intrinsic electrical resistance of the nanoobject. **3) Four** Au NW electrodes to measure the electrical resistance of the nanoobject without contact resistance. Topologies 2 and 3 provide stand-alone devices in the sense that they are fully independent of the AFM used to build them. Consequently, they can be taken to other experimental setups for further characterization. In total, we are reporting up to 10 different circuits described in the main article and the SI.

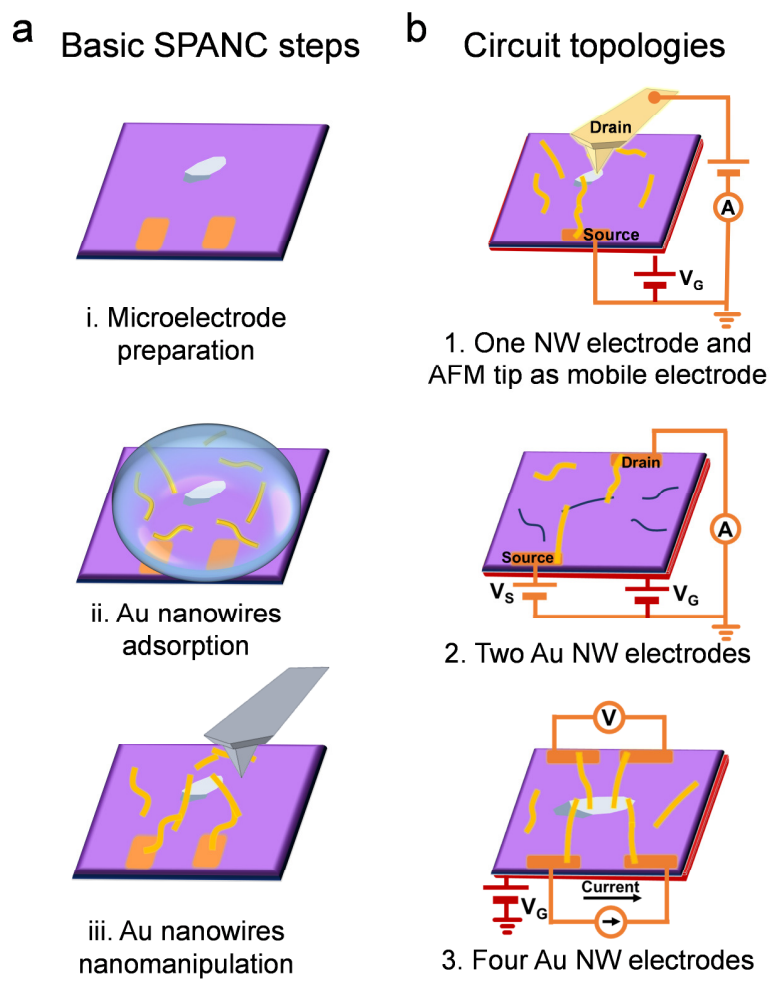

**Fig. 1.** Basic SPANC steps and circuit topologies. (a) Preparation steps for SPANC. The different panels show schematics of the steps. Top: microelectrode fabrication. Middle: Au NWs adsorption on the substrate. Bottom: AFM manipulation to create the nanolectrodes. (b) Schematics of the different contact configurations used in this work. Top: one Au NW electrode and a conductive tip as a second mobile electrode. Middle: two Au NW electrodes. Bottom: four Au NW electrodes.

**One NW electrode and a conductive AFM tip as a second mobile electrode setup. Electrical characterization of few layer antimonene.** The origin of SPANC goes back to the characterization of the electrical transport properties of few layer antimonene flakes.[18] To this end, we followed the standard procedures using EBL to fabricate nanocircuits (see SI2). Fig. S2-1 in the SI portrays a few layer antimonene flake with four electrical nanoelectrodes made by EBL. For unknown reasons, probably due to antimonene instability to the chemicals used in the EBL procedure, we could never measure electrical current through the flakes. After many attempts, we envisioned SPANC as an alternative technique for electrical connectivity.

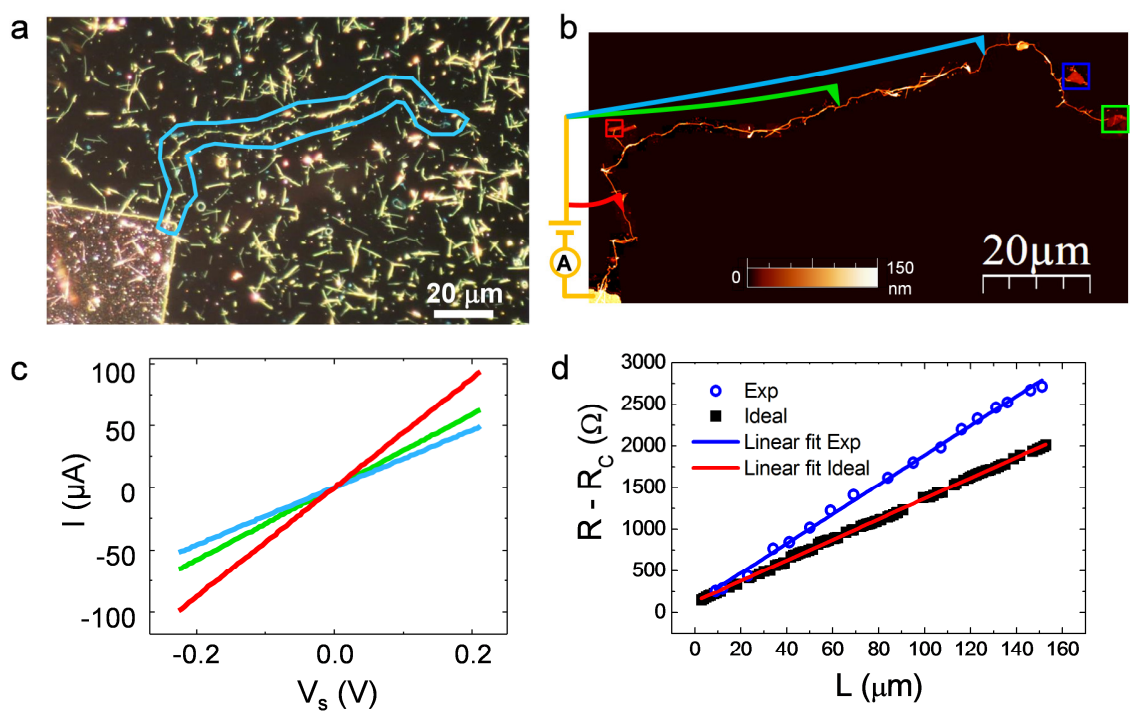

**Fig. 2.** Au NW paths characterization. (a) Dark field optical microscopy image showing a microelectrode (bottom left) and a random distribution of nanowires. The blue line encircles a nanowires path that we made by AFM nanomanipulation. (b) Atomic force microscopy topographic image showing the nanowire path. For the sake of clarity, we suppressed the rest of the nanowires. The path connects three thin antimonene flakes (enclosed inside squares) to the microelectrode at the bottom left. We used a conductive AFM tip as a second mobile electrode. (c) Current *vs.* voltage characteristics acquired along the path at the different locations indicated by the schematic cantilevers in (b). (d) Resistance *vs.* length plot (blue circles) measured along the nanowires path. The black squares represent the ideal resistance of this path assuming that the resistivity of the nanowires is the same as the bulk gold. The difference between both lines yields the nanowire-nanowire contact resistance.

Fig. 2a shows a dark field optical microscope image where a 150 μm long path formed by 93 Au NWs (enclosed in cyan) connects three different few layer antimonene flakes to a gold microelectrode. Coexisting with these NWs conforming the path, there are many other NWs not used to build the circuit. The nominal dimensions of the NWs are ~7 μm in length with ~50 nm diameter.[19] For the sake of clarity, we depict in Fig. 2b the corresponding AFM topographic image but now suppressing the NWs that do not participate in the circuit. The electric circuit in this case corresponds to the one portrayed in Fig. 1b with just one fixed microelectrode and a conductive tip acting as a second mobile electrode.[15] We can plot the resistance of the path as a function of its length (Fig. 2d) by acquiring current *vs.* voltage (*IV*) characteristics at different locations along the path (see schematics in Fig. 2b and three different *IV*s in Fig. 2c). From this plot we can quantify the electrical resistance of the NW junctions. We compare the experimental resistance *vs.* length of the path with its ideal resistance, obtained by calculating the volume of each NW (from diameter and length values from AFM imaging) and considering that the resistivity of a single NW is that of bulk gold.[20] The difference between the ideal resistance at the end of the path and the measured one is about 800 Ω. Hence, as we have 93 NW junctions, we end up with a figure of ~9 Ω/junction. Studies on the conductivity of metallic thin films and wires conclude that grain boundary reflections rather than surface scattering are the dominant contributions to the resistivity of Au NWs.[21, 22] Reference [20] demonstrates that nanowires with the same geometry and crystallinity as the ones used here present the same resistivity as bulk gold. Then, we conclude that the obtained resistance per junction can be attributed mainly to the formation of just a single grain boundary in each of the junctions.[23] This low resistance is indeed excellent for nanocircuitry.

Next, we characterize the transport properties of the three few layer antimonene flakes in Fig. 3. For the sake of clarity, we also enclosed these flakes in Fig. 2b (red, blue and green rectangles). The thicknesses of the flakes range between 4 and 21 nm. Gold nanowires are in contact with the flakes on their left sides. Fig. 3d depicts 4 *IV* curves corresponding to the triangles superimposed across the flake. The slope of the lines increases as the distance between the conductive tip and the NW decreases. The 5[th] curve corresponds to and *IV* taken on the nanowire, showing the lowest resistance. The circuit included a 2 kΩ protection resistor, a common procedure in nanocircuits, which is reflected as an additional resistance in the *IV*s. Fig. 3e presents a plot showing different resistance *vs.* length (*RL*) plots acquired on the flakes by performing *IV* characteristics on them. We used two different AFM conductive tips, finding similar results (notice the same slope in the *RL*s acquired with different tips on the same flake and only an offset between them, coming from a different contact resistance). From the resistance *vs.* length plots we can estimate a sheet resistance for antimonene flakes of $\rho_{2D}$ = 1300 ± 400 Ω/□. Electrical characterization of thin antimony crystals grown by van der Waals epitaxy with thicknesses of 30-50 nm has been reported,[24] but to the best of our knowledge, this is the first electrical characterization of antimonene flakes so thin, with thicknesses within a regime where topological effects have been theoretically predicted.[25]

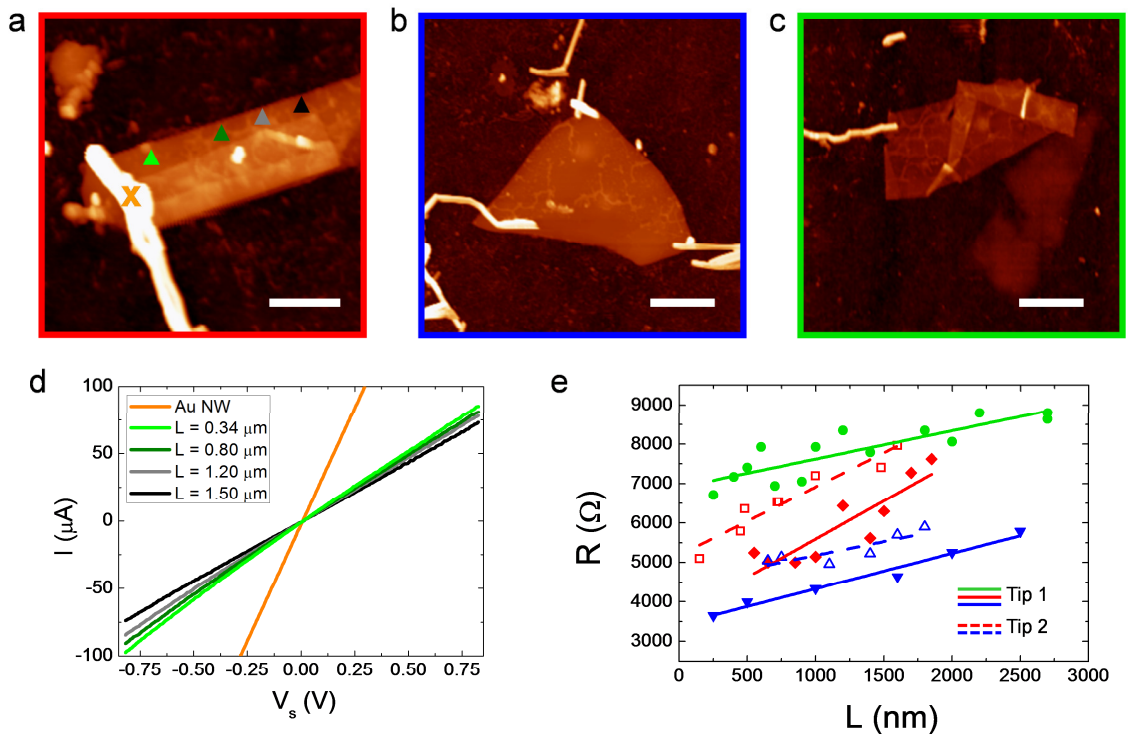

**Fig. 3.** (a-c) Close up images of the antimonene flakes enclosed in the red, blue and green squares in accordance with Fig. 2b. Scale bars: 600 nm, 1.2 μm and 1.2 μm respectively. (d) *IV* curves taken in the "X" (Au NW electrode) and the triangles marked in (a). (e) Resistance *vs.* length characteristics taken on the few layer antimonene flakes, each color representing a flake in accordance to (a-c). Each symbol corresponds to an *IV* curve from where we calculate the resistance. Lines are linear fits. Filled symbols and solid lines correspond to data acquired with one tip, whereas empty symbols and dashed lines correspond to a different tip.

For the sake of completeness, we validate this electrode configuration (see SI3 and Fig. S3-1) using a few layer graphene flake as fiducial sample obtaining a sheet resistivity $\rho_{2D}$ = 670 ± 60 $\Omega/\square$, in good agreement with values reported through different electrical contact schemes.[2, 26, 27] Importantly, the total time required for these measurements was less than 4 hours, discounting both the preparation of the flakes and the evaporation of the microelectrodes.

**Two NW electrode setup**. Fig. 4 portrays two different experiments carried out using two NW electrodes fabricated *via* SPANC. Fig. 4a presents an AFM topography showing a gold microelectrode (left), a graphene nanoribbon (enclosed in the figure) and two NW paths connecting the nanoribbon with the microelectrodes (the upper electrode is on the top right side, outside the image). After creating the electrical contacts with Au NWs, we placed the sample in an electrical transport measurement setup inside a vacuum chamber allowing electrical characterization between 96 and 373 K. Figs. 4b and 4c show the dependence of the electrical current with the drain-source bias voltage and with a global backgate voltage applied through the 300 nm $SiO_2$ layer. With the sample in vacuum we monitored the circuit for eleven days to demonstrate the SPANC-made circuits endurance to temperature and pressure variations.

Fig. 4d shows an AFM topography of a multiwalled carbon nanotube connected to two Au NW paths. By reconfiguring the position of the upper Au NW (see sequence in Fig. S3-2), bringing it down to the vicinity of the lower Au NW electrode, we can obtain the dependence of the electrical resistance of the nanotube on the distance between electrodes (*L)* (Fig. 4e). We estimate a contact resistance of ~20 k$\Omega$, which is an excellent figure for a carbon nanotube contact resistance.[15] This experiment provides an example of electrode reconfiguration, *i.e.* the position of the Au NWs working as electrodes can be varied in a simple and quick manner. Finally, Fig. S3-3 shows the dependence of the drain-source current on the gate voltage.

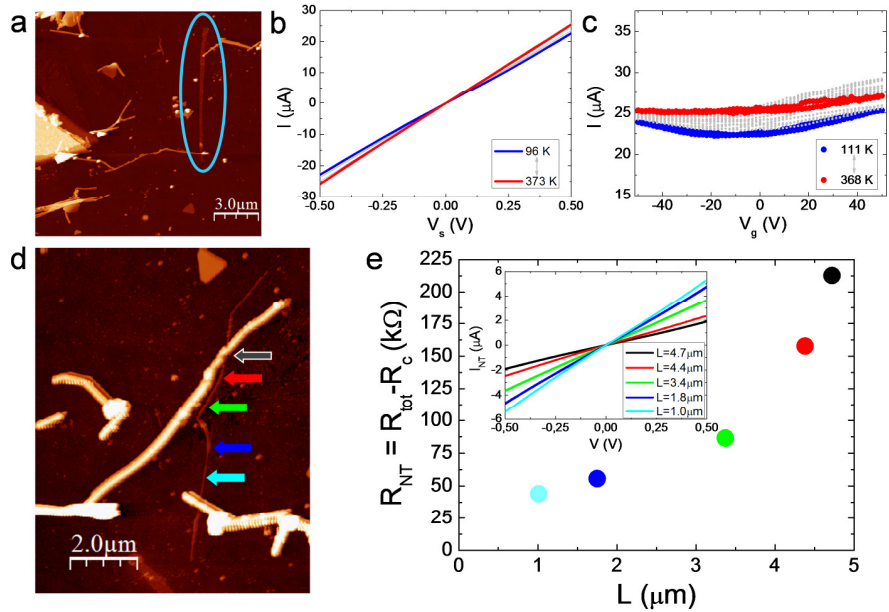

**Fig. 4.** Two Au NW electrode setup experiments. (a) AFM topography showing a graphene nanoribbon connected to two nanowires paths. (b) Drain-source current as a function of the bias voltage at gate voltage zero and different temperatures between 96 and 373 K for the nanoribbon shown in (a) in the stand-alone configuration. (c) Current *vs*. gate voltage dependence in the same conditions as in (b) (*i.e.* $V_s$ = 0.5 V). (d) AFM image of a multiwalled carbon nanotube connected to two nanowires paths. The arrows mark the locations where *IV* curves were acquired by repositioning the upper nanowire. (e) Resistance *vs*. distance plot obtained by approaching the upper gold nanowire to the lower one. The color dots correspond to the positions marked by the corresponding color arrows in (d). The inset shows the acquired *IV* characteristics.

**Four NW electrode setup**. Fig. 5 summarizes the main results of a four contacts device carried out to study the transport properties of a few layer graphene flake. Although the fabrication procedure and the type of device of this setup are similar to the two NW electrodes scheme, this Kelvin sensing topology enables the experimental measurement of the resistance of a nanoobject without contact resistances. This configuration is the one required if the contact resistance is higher than the intrinsic nanoobject resistance. Fig. 5a presents an AFM topography showing the flake and the four Au NW paths (encircled in the figure) used. The path in the bottom left corner (P2) is a single Au NW electrically connecting the flake to the microelectrode at the bottom of the image. The other Au NW paths (P1, P3 and P4) connect, in the same way, the flake (at different points) with three different microelectrodes (outside the imaged area). The red line in Fig. 5b shows the dependence of the electrical current with the drain-source voltage in a two-contact configuration using paths P2 and P4. For comparison, the steeper blue line represents the dependence of the current *vs.* voltage in the four-contact topology. The slope difference between both lines reflects the presence of a high contact resistance of ~20 kΩ in the two contacts scheme measurement. Nevertheless, by using the four-wire sensing setup the measured current is not affected by this contact resistance. Fig. 5c portrays a plot of the flake electrical resistance as a function of the distance. We carried out this measurement by leaving path P4 unconnected and using a conductive AFM tip as a fourth mobile electrode (always placed between P2 and P3). The advantage with respect to the typical conductive AFM measurements (with a two-electrode topology) is the absence of contact resistance. Finally, Fig. 5d shows the current *vs.* voltage variation in a four-terminal configuration, but now acquired in the stand-alone setup (the P4 Au NW path was used again), with the sample inserted in a vacuum chamber at variable temperature from 100 K to room temperature.

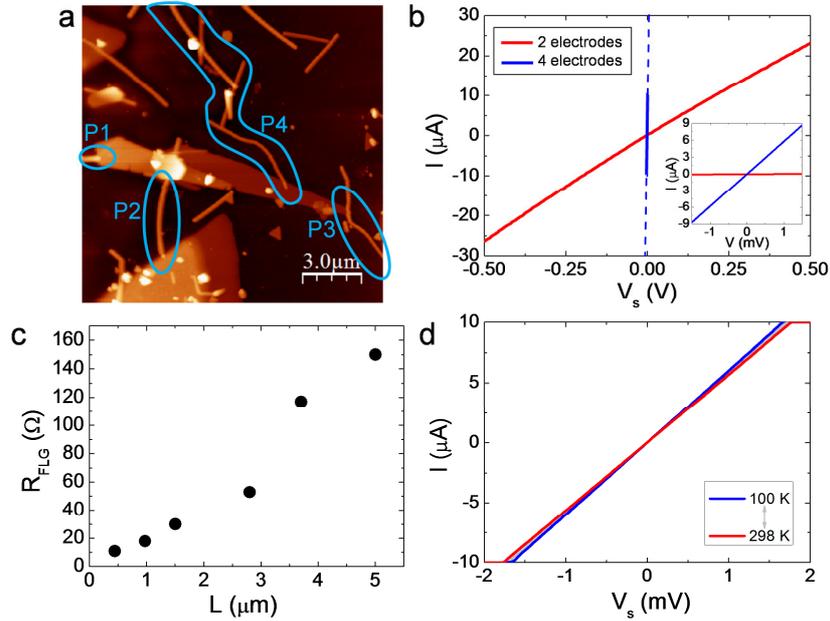

**Fig. 5.** Four Au NW electrode setup experiments. (a) AFM topography of a few layer graphene flake connected to four nanowires paths (P1, P2, P3 and P4). In the lower side of the image one of the four microelectrodes is visible. (b) Current *vs*. voltage curves taken in the two-electrode configuration (red line) and in the four electrode configuration (blue line). The inset corresponds to a zoom in of the same plot at low voltage. (c) Resistance *vs*. distance plot obtained by replacing electrode P4 by a metallized AFM tip used as mobile electrode. (d) Drain-source current *vs*. voltage characteristics in the stand-alone setup between two different temperatures: 100 and 298 K.

**Other relevant nanocircuits.** The examples discussed above deal with nanoobjects with dimensions similar to those connected using conventional methods to fabricate nanocircuits. In this section we present two additional examples which go beyond that spatial resolution and illustrate further the possibilities of SPANC.

In order to show the limits of the technique we proceed to evaporate gold microelectrodes on a conventional $SiO_2$ substrate followed by Au NWs deposition. Then we deposit gold nanoparticles (Au NPs), with an average diameter of 15 nm, to form a mix-sample that contains both NWs and NPs. By comparing the images before and after the deposition of the nanoparticles (see SI4 and Fig. S4-1) it is straightforward to identify the nanoparticles. We formed two continuous NW paths which connect two gold nanoparticles in parallel, as shown in Fig. 6a. The heights of the bottom and top NPs are ~20 and ~10 nm, respectively. The top particle was in contact with the NWs using a third ~25 nm NP that facilitates the connection between the two NWs and the smallest particle. Fig. 6b shows the final *IV* characteristic of the device. The resistance of the *IV* suggests that the contact resistances between the NWs and the NPs are low. The inset depicts the schematics of the final assembly, obtained after a series of manipulations shown in Fig. 6c. This figure illustrates the high control of the nanoelectrode positioning that can be achieved with SPANC (see also Fig. S4-2 in the SI for details of the capture of a gold nanoparticle with a NW for the device assembly). This experiment allows us to conclude that the size of the smallest nanoobject we can contact with SPANC at this stage is ~10 nm. Additionally, it is important to remark that the total elapsed time for this experiment was 3-4 days.

Finally, we introduce a preliminary experiment towards molecular electronics, which would be unfeasible using conventional nanoelectrode fabrication techniques. Molecular electronics is a field dominated by experimental methods such as: scanning probe microscopy,[28] molecular break junctions (MBJ)[29, 30] or molecular trapping,[31] to mention some of them. The basic idea is to form atomic size gaps that are "filled" with nanoobjects, very often in a random manner.[32] By inducing mechanical stress, the gap width can be readjusted creating new molecular junctions. The current flowing through the junctions so formed is measured and analyzed by statistical methods (histograms and cluster analysis) that allow quantifying the conductance of the molecules.[33]

Our experiments started with two gold nanowires, each connected to two separated gold microelectrodes. Then we used the AFM tip to approach the nanowires as much as possible (Fig. 6d) but avoiding cold welding (we used a sharp tip to check the gap width, see SI5and Fig. S5-1). Once the nanowires were so prepared, no current was observed in this configuration applying voltages up to 4.5 V. We first induced an oxygen plasma to remove the CTAB surfactant and then we deposited 1,4- benzenedithiol (BDT) molecules on the device (see details of the sample preparation in Methods). Therefore, a ~1 nm thick layer of BDT was covering the electrode junction area, as we assume by a similar deposition and characterization by ellipsometry on a flat gold substrate. Upon the BDT layer deposition, we imaged the sample again observing no variations in the nanowires position and we confirmed the absence of cold welding. We then achieved the onset of current applying bias voltages up to 5V (see SI5 and Fig. S5-2 for details). *IV*s at lower bias voltages produced the same conductance in all of them (see Fig. S5-2). We checked the stability of the contacts so formed for more than one hour, yielding a quite stable conductance of about 0.06$G_0$ (Fig. 6e). This conductance suggests NW-NW gaps of ~1 nm (about the length of a BDT molecule), in agreement with transmission electron microscopy imaging of two adjacent NWs in another device (see Fig. S5-4). The conductance of BDT molecules depends critically on the experimental setup and different groups reported values ranging from ~$10^{-4}G_0$ to ~0.5$G_0$,[34-36] in good agreement with our experimental results. Additionally, the spectral analysis of conductance temporal traces (see Fig. S5-5) suggests the existence of a Random Telegraph Signal superimposed on the background noise.[37] This noise distribution is characteristic of nanojunctions that in many cases involve just a few molecules.[38]

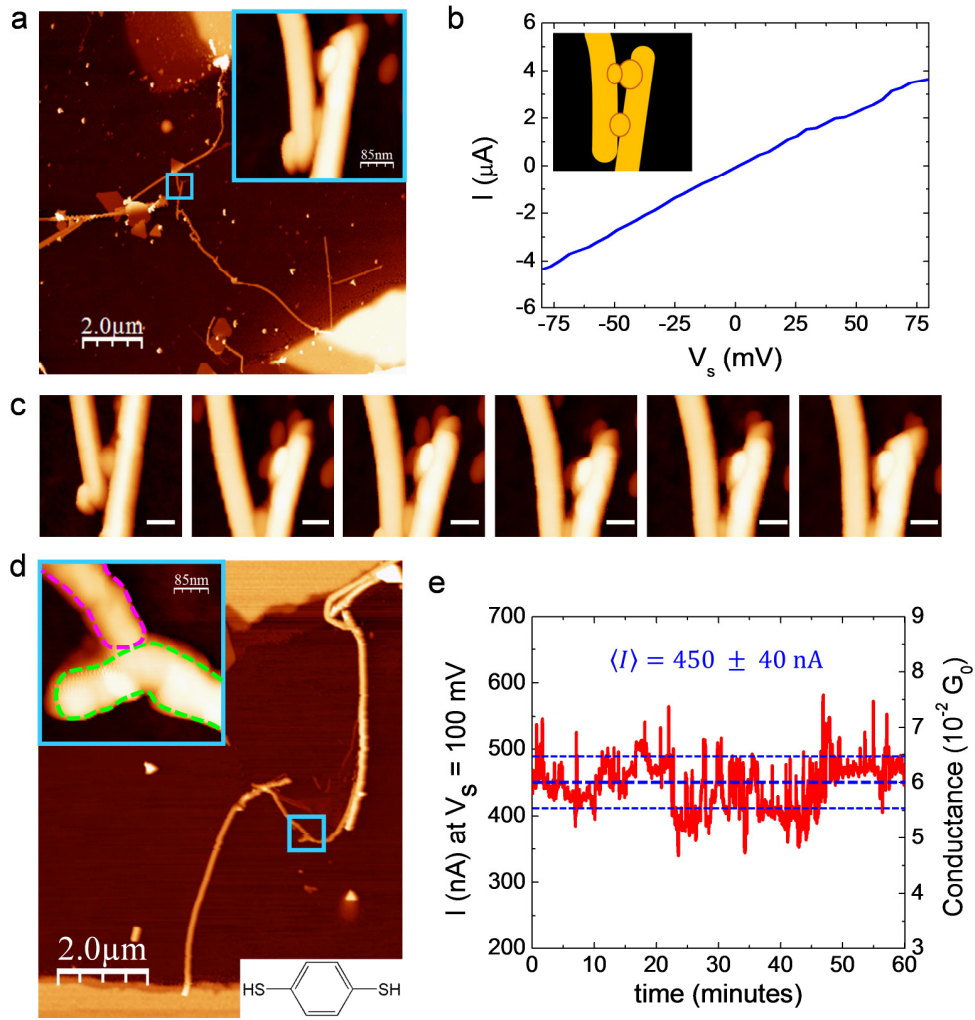

**Fig. 6.** Other relevant nanocircuits. (a) AFM topographic image of two Au NW electrodes connecting two gold nanoparticles in parallel. The inset corresponds to a zoom in the region enclosed by the blue square in the main figure. (b) Current *vs*. voltage characteristic of the circuit in (a). The inset is a schematic view of the final device as shown in (a). (c) Sequence of manipulations of the Au NWs to bring the top Au NP in contact with them. Scale bars: 70 nm. (d) AFM topographic image of two Au NW electrodes brought into close proximity. The inset on the top left corner corresponds to a zoom in the region enclosed by the blue square in the main figure. The dotted lines show the edges of the NWs. Despite no noticeable gap between the NWs within tip resolution, the circuit exhibited a resistance higher than 1 TΩ, the highest resistance that we could measure with our experimental set up. The inset in the bottom right corner shows the structure of the 1,4- benzenedithiol molecules that were deposited on this device. (e) Current/Conductance variation through the NW-BDT-NW contact along one hour when a bias voltage of 100 mV was applied. The mean current ± standard deviation values are shown graphically and numerically.

In the molecular electronics experiment described above we cannot completely discard some complex combination of tunneling and molecule transport, or some contamination such as CTAB rests. This experiment allows electrical measurement through a molecule at the time that an image of the interesting region is acquired. This experiment could be improved in many different ways, as for instance inserting the AFM in a vacuum chamber with a controlled atmosphere of the molecule of interest. Nevertheless, what we would like to remark is the stability of the SPANC configuration *vs.* STM or break junctions. Whereas these two techniques are three-dimensional setups, in SPANC the molecular junction is restricted to the substrate plane. Due to this extra degree of freedom, STM and MBJ are intrinsically less stable than SPANC. Using STM and MBJ measurements longer than one minute are a challenge. The standard remedy to increase stability consists in working at cryogenic temperatures, but then the experimental setups become much more complicated.[28, 38] On the contrary, SPANC presents a two dimensional configuration that inherently stabilizes the system and the time stability goes up to more than an hour (Fig. 6e).

Finally, we confirmed the possibility of using SPANC in other insulating substrates such as mica. The fabrication of circuits on this substrate revealed no significant differences respect to silicon dioxide (see SI6). We also tested the influence of the humidity. To this end, we used an AFM inside a stainless steel vacuum chamber where we could vary the relative humidity from negligible humidity (total pressure less than $10^{-2}$ mbar) up to 80 ± 5 % RH. Again, our experiments did not show any significant difference respect to circuit fabrication in air ambient conditions (see SI7).

**CONCLUSIONS**

So far, the use of highly conductive nanostructures, as those fabricated by SPANC, was a long-standing goal of circuit miniaturization.[17, 39, 40] In this work, we have demonstrated that SPANC is a cost-effective, versatile and robust technique that provides nanocircuitry for stand-alone devices, allowing electrical connections in nanoobjects as small as ~10 nm. We have proved that the fabrication of circuits can be carried out on different substrates and humidity conditions. The requirements for cantilevers are moderate (one single tip can be used for the manipulation of many wires) and the force range used for manipulation is in the order of 100 nN. Unlike conventional techniques, SPANC also allows circuit reconfiguration in a straightforward and quick manner. Besides, SPANC requirements and preparation are simple and clean as it uses drop casting of a water or alcohol (ethanol, isopropanol, methanol, etc.) suspension of nanowires. Hence, SPANC is compatible with delicate materials. It is not difficult to envision combined nanocircuitry of EBL and SPANC for the electrical characterization of nanoobjects of a few nanometers and materials that do not support the chemical agents required for EBL, as emerging organics, organometallics and even biomolecules, among others. In the future, we aim at introducing a broader catalogue of possibilities that will include the use of longer/shorter and thicker/thinner nanowires made of different materials: silver, nickel, cobalt, platinum, etc. They can be used in combination with other sophisticated elements such as carbon nanotubes to improve connectivity to molecules. In some cases, the throughput of SPANC is even better than EBL. Moreover, in the future SPANC could be improved by using parallel processing[41] and image recognition combined with the automation of the tip trajectories for the NWs manipulation and assembly. In summary, albeit SPANC is in its very infancy and the results presented herein give a highly limited picture of its many possibilities, we consider SPANC is now a reality with a promising future ahead of it.

**METHODS**

*Nanowires*. We used two different sources of nanowires. We assembled the circuits shown in Figs. 2, 3 and Fig. S3-1 with commercial gold nanowires from Nanopartz Inc. (www.nanopartz.com). For the rest of the experiments, we prepared home-synthesized gold nanowires following a modification of the literature method,[42] based on the well-established three-step protocol for nanorods synthesis published by Murphy *et al.*;[43] its synthesis first involves the preparation of aqueous gold seed nanoparticle of 3-5 nm by reduction of the $HAuCl_4$ solution by $NaBH_4$ in the presence of sodium citrate. For the growth of nanowires, the seed particles are added to a growth solution containing a mixture of $HAuCl_2$ ($Au^+$) which was reduced from $Au^{3+}$ salt by a weak reducing agent as ascorbic acid in the presence of a structure directing agent as CTAB. The growth was allowed to continue for at least 48h. See Supporting

Note SI8 for additional details and Fig. S8-1 for scanning electron images of the synthesized nanowires.

*Nanowires preparation and deposition*. We first placed the dispersion containing the Au NWs on a vortex mixer for 15 seconds, then we heated and sonicated a 200 μl aliquot in an ultrasound bath (30°C, 37 kHz-380 W) for 2-5 minutes to resuspend aggregated NWs. Afterwards, we cast a 10 μl drop on the substrate and left it for 10 minutes at a temperature of almost 30°C. After that time, we washed the substrate with deionized (DI) water at this same temperature and dried it with a $N_2$ gas flow. We performed both washing and drying processes very smoothly to avoid dragging of the NWs. We carried out all the steps at ~30°C to avoid aggregation of CTAB, which tends to form crystals. We then inspected the NW substrate with an optical microscope in dark field mode to check the density of NWs and repeated the drop-casting step (using the same aliquot) until a concentration of around 1 NW per 5 μm$^2$ was achieved. This usually requires several repetitions. Just before each drop casting step, the aliquot was agitated with the vortex mixer to prevent the NWs from depositing at the bottom of the vial.

*Optical microscopy*. We used a Zeiss Axiovert microscope equipped with x5, x10, x20, x50 and x100 set of lenses. We routinely took dark field images to check the density and spatial distribution of the NWs.

*Atomic force microscopy imaging*. We carried out AFM measurements using a Cervantes Fullmode AFM from Nanotec Electronica SL. We employed WSxM software (www.wsxm.es) both for data acquisition and image processing.[44,45] We acquired topography images in amplitude modulation mode using PPP-FM cantilevers from Nanosensors, with nominal resonance frequency of 75 kHz and spring constant of 2.8 Nm$^{-1}$. We used these same cantilevers to manipulate the NWs. We checked the electrical resistance of the NW paths using conductive diamond coated tips (Budget Sensors All-In-On-DD). For the acquisition of data of Figs. 2, 3 and Fig. 5c we used CrPt coated tips (Budget Sensors ElectriMulti75-G) to improve the electrical contact between the sample and the tip. Metal coated tips can be also employed to check the resistance of the NW nanoelectrodes, but we have observed that they present a high adhesion with the NWs which can result in unwanted NW extraction.

*Thermal evaporation*. We subsequently deposited 10-15 nm of Cr followed by a thicker layer of gold (~100 nm) on the substrates (300 nm oxide silicon grown thermally on a highly-doped Si (111)).

*Evaporation masks*. We used *ad hoc* stencil masks made of nickel for thermal evaporation. The masks offer several topologies for the deposition of microelectrodes (see SI9 and Fig. S9-1 for additional details) and are now commercially available from Gilder Grids (www.gildergrids.co.uk). We will be glad shipping mask samples upon requirement.

*Nanowires manipulation*. We moved the NWs by using the lithography option of WSxM.[44] We performed the motion of the NWs to assembly the paths by moving the tip in contact mode along predefined trajectories[46] (see Supporting Note SI1). The normal force required during the nanomanipulation ranges from 60 to 300 nN. For the coarse NW manipulations we navigated areas of ~100-200 $\mu m^2$, going down to ~5-10 $\mu m^2$ for the more delicate assemblies. Although AFM tips can wear during manipulation, which translates into a wider apparent width of the NWs, this is not an issue for the majority of the NW assemblies. To have an idea, during the formation of the 150 µm long path in Fig. 2, we were able to cover the assembly along distances of at least ~40 µm with the same tip. This is indeed similar to the total distance assembled for the four-electrode device in Fig. 5. Drift was not relevant on our manipulations, as can be observed in Fig. 6c, where the precise positioning control of the NWs can be observed.

*1,4-benzenedithiol (BDT) device preparation*. Once the Au NW electrodes were arranged as close as possible but without being in electrical contact, we inserted the sample in a vacuum chamber (base pressure $10^{-6}$ mbar). Firstly, we induced an oxygen plasma in order to remove the CTAB surfactant. Secondly, we introduced a BDT partial pressure of $5 \times 10^{-4}$ mbar for 30 seconds. Previously, we had checked by ellipsometry that this very same procedure applied to a gold thin film produced a molecular layer of ~0.9 nm, compatible with the size of a BDT molecule.

**ACKNOWLEDGMENT**

We would like to thank Eva Durán, Dr. Carmen Montoro, Diego A. Aldave, Dr. Eduardo Lee, Dr. Adriana Gil and Dr. Laura Fumagalli for their help in the preparation of some samples and for insightful discussions. We acknowledge financial support through the "María de Maeztu" Programme for Units of Excellence in R&D (MDM-2014-0377) and from projects MAT2016-77608-C3-1-P, and MAT2016-77608-C3-3-P, MAD2D-CM and MAT2013-46753-C2-2-P that includes Miriam Moreno-Moreno's FPI fellowship. Ramon Areces foundation is acknowledged for financial support.

**REFERENCES**

1. McCord, M. A.; Rooks, M. J., Handbook of Microlithography, Micromachining and Microfabrication. In *Handbook of Microlithography, Micromachining and Microfabrication*, Rai-Choudhury, P., Ed. 1997; Vol. 1: Microlithography.

2. Novoselov, K. S.; Geim, A. K.; Morozov, S. V.; Jiang, D.; Zhang, Y.; Dubonos, S. V.; Grigorieva, I. V.; Firsov, A. A. Electric field effect in atomically thin carbon films. *Science* **2004**, *306*, 666-669.

3. Tans, S. J.; Verschueren, A. R. M.; Dekker, C. Room-temperature transistor based on a single carbon nanotube. *Nature* **1998**, *393*, 49-52.

4. Hatzakis, M. Electron resists for microcircuit and mask production. *J. Electrochem. Soc.* **1969**, *116*, 1033-1037.

5. Manfrinato, V. R.; Zhang, L.; Su, D.; Duan, H.; Hobbs, R. G.; Stach, E. A.; Berggren, K. K. Resolution limits of electron-beam lithography toward the atomic scale. *Nano Lett.* **2013**, *13*, 1555-1558.

6. Binnig, G.; Quate, C. F.; Gerber, C. Atomic force microscope. *Phys. Rev. Lett.* **1986**, *56*, 930-933.

7. Gross, L.; Mohn, F.; Moll, N.; Liljeroth, P.; Meyer, G. The chemical structure of a molecule resolved by atomic force microscopy. *Science* **2009**, *325*, 1110-1114.

8. Piner, R. D.; Zhu, J.; Xu, F.; Hong, S. H.; Mirkin, C. A. "Dip-pen" nanolithography. *Science* **1999**, *283*, 661-663.

9. Day, H. C.; Allee, D. R. Selective area oxidation of silicon with a scanning force microscope. *Appl. Phys. Lett.* **1993**, *62*, 2691-2693.

10. Garcia, R.; Calleja, M.; Perez-Murano, F. Local oxidation of silicon surfaces by dynamic force microscopy: Nanofabrication and water bridge formation. *Appl. Phys. Lett.* **1998**, *72*, 2295-2297.

11. Rawlings, C.; Ryu, Y. K.; Ruegg, M.; Lassaline, N.; Schwemmer, C.; Duerig, U.; Knoll, A. W.; Durrani, Z.; Wang, C.; Liu, D. X.; Jones, M. E. Fast turnaround fabrication of silicon point-contact quantum-dot transistors using combined thermal scanning probe lithography and laser writing. *Nanotechnology* **2018,** *29*, 505302.

12. Durrani, Z.; Jones, M.; Abualnaja, F.; Wang, C.; Kaestner, M.; Lenk, S.; Lenk, C.; Rangelow, I. W.; Andreev, A. Room-temperature single dopant atom quantum dot transistors in silicon, formed by field-emission scanning probe lithography. *J. Appl. Phys.* **2018,** *124*, 144502.

13. Calleja, M.; Tello, M.; Anguita, J.; Garcia, F.; Garcia, R. Fabrication of gold nanowires on insulating substrates by field-induced mass transport. *Appl. Phys. Lett.* **2001**, *79*, 2471-2473.

14. Bachtold, A.; Henny, M.; Terrier, C.; Strunk, C.; Schonenberger, C.; Salvetat, J. P.; Bonard, J. M.; Forro, L. Contacting carbon nanotubes selectively with low-ohmic contacts for four-probe electric measurements. *Appl. Phys. Lett.* **1998**, *73*, 274-276.

15. Gomez-Navarro, C.; De Pablo, P. J.; Gomez-Herrero, J.; Biel, B.; Garcia-Vidal, F. J.; Rubio, A.; Flores, F. Tuning the conductance of single-walled carbon nanotubes by ion irradiation in the Anderson localization regime. *Nat. Mater.* **2005**, *4*, 534-539.

16. Thelander, C.; Samuelson, L. AFM manipulation of carbon nanotubes: realization of ultra-fine nanoelectrodes. *Nanotechnology* **2002**, *13*, 108-113.

17. Lu, Y.; Huang, J. Y.; Wang, C.; Sun, S. H.; Lou, J. Cold welding of ultrathin gold nanowires. *Nat. Nanotechnol.* **2010**, *5*, 218-224.

18. Ares, P.; Aguilar-Galindo, F.; Rodriguez-San-Miguel, D.; Aldave, D. A.; Diaz-Tendero, S.; Alcami, M.; Martin, F.; Gomez-Herrero, J.; Zamora, F. Mechanical isolation of highly stable antimonene under ambient conditions. *Adv. Mater.* **2016**, *28*, 6332-6336.

19. Kim, F.; Sohn, K.; Wu, J.; Huang, J. Chemical synthesis of gold nanowires in acidic solutions. *J. Am. Chem. Soc.* **2008**, *130*, 14442-14443.

20. Critchley, K.; Khanal, B. P.; Gorzny, M. L.; Vigderman, L.; Evans, S. D.; Zubarev, E. R.; Kotov, N. A. Near-bulk conductivity of gold nanowires as nanoscale interconnects and the role of atomically smooth interface. *Adv. Mater.* **2010**, *22*, 2338-2342.

21. Steinhogl, W.; Steinlesberger, G.; Perrin, M.; Scheinbacher, G.; Schindler, G.; Traving, M.; Engelhardt, M. Tungsten interconnects in the nano-scale regime. *Microelectron. Eng.* **2005**, *82*, 266-272.

22. Wu, W.; Brongersma, S. H.; Van Hove, M.; Maex, K. Influence of surface and grain-boundary scattering on the resistivity of copper in reduced dimensions. *Appl. Phys. Lett.* **2004**, *84*, 2838-2840.

23. Bietsch, A.; Michel, B. Size and grain-boundary effects of a gold nanowire measured by conducting atomic force microscopy. *Appl. Phys. Lett.* **2002**, *80*, 3346-3348.

24. Ji, J.; Song, X.; Liu, J.; Yan, Z.; Huo, C.; Zhang, S.; Su, M.; Liao, L.; Wang, W.; Ni, Z.; Hao, Y.; Zeng, H. Two-dimensional antimonene single crystals grown by van der Waals epitaxy. *Nat. Commun.* **2016**, *7*, 13352.

25. Hsieh, D.; Xia, Y.; Wray, L.; Qian, D.; Pal, A.; Dil, J. H.; Osterwalder, J.; Meier, F.; Bihlmayer, G.; Kane, C. L.; Hor, Y. S.; Cava, R. J.; Hasan, M. Z. Observation of unconventional quantum spin textures in topological insulators. *Science* **2009**, *323*, 919-922.

26. Han, M. Y.; Oezyilmaz, B.; Zhang, Y.; Kim, P. Energy band-gap engineering of graphene nanoribbons. *Phys. Rev. Lett.* **2007**, *98*, 206805.

27. Moreno-Moreno, M.; Castellanos-Gomez, A.; Rubio-Bollinger, G.; Gomez-Herrero, J.; Agrait, N. Ultralong Natural Graphene Nanoribbons and Their Electrical Conductivity. *Small* **2009**, *5*, 924-927.

28. Pascual, J. I.; Mendez, J.; Gomezherrero, J.; Baro, A. M.; Garcia, N.; Binh, V. T. Quantum contact in gold nanostructures by scanning-tunneling-microscopy. *Phys. Rev. Lett.* **1993**, *71*, 1852-1855.

29. Agrait, N.; Yeyati, A. L.; van Ruitenbeek, J. M. Quantum properties of atomic-sized conductors. *Phys. Rep.* **2003**, *377*, 81-279.

30. vanRuitenbeek, J. M.; Alvarez, A.; Pineyro, I.; Grahmann, C.; Joyez, P.; Devoret, M. H.; Esteve, D.; Urbina, C. Adjustable nanofabricated atomic size contacts. *Rev. Sci. Instrum.* **1996**, *67*, 108-111.

31. Porath, D.; Bezryadin, A.; de Vries, S.; Dekker, C. Direct measurement of electrical transport through DNA molecules. *Nature* **2000**, *403*, 635-638.

32. Park, H.; Lim, A. K. L.; Alivisatos, A. P.; Park, J.; McEuen, P. L. *Appl. Phys. Lett.* **1999**, *75*, 301-303.

33. Cuevas, J. C.; Scheer, E., M*olecular electronics: an introduction to theory and experiment*. 2010.

34. Kim, Y.; Pietsch, T.; Erbe, A.; Belzig, W.; Scheer, E. Benzenedithiol: A broad-range single-channel molecular conductor. *Nano Lett.* **2011**, *11*, 3734-3738.

35. Xiao, X. Y.; Xu, B. Q.; Tao, N. J. Measurement of single molecule conductance: Benzenedithiol and benzenedimethanethiol. *Nano Lett.* **2004**, *4*, 267-271.

36. Yamauchi, K.; Kurokawa, S.; Sakai, A. Admittance of Au/1,4-benzenedithiol/Au single-molecule junctions. *Appl. Phys. Lett.* **2012**, *101*, 253510.

37. Yuzhelevski, Y.; Yuzhelevski, M.; Jung, G. Random telegraph noise analysis in time domain. *Rev. Sci. Instrum.* **2000**, *71*, 1681-1688.

38. Brunner, J.; Teresa Gonzalez, M.; Schoenenberger, C.; Calame, M. Random telegraph signals in molecular junctions. *J. Phys.: Condens. Matter* **2014**, *26*, 474202.

39. Stern, A.; Eidelshtein, G.; Zhuravel, R.; Livshits, G. I.; Rotem, D.; Kotlyar, A.; Porath, D. Highly conductive thin uniform gold-coated DNA nanowires. *Adv. Mater.* **2018**, *30*, 1800433.

40. Pate, J.; Zamora, F.; Watson, S. M. D.; Wright, N. G.; Horrocks, B. R.; Houlton, A. Solution-based DNA-templating of sub-10 nm conductive copper nanowires. *J. Mater. Chem. C* **2014**, *2*, 9265-9273.

41. Vettiger, P.; Despont, M.; Drechsler, U.; Durig, U.; Haberle, W.; Lutwyche, M. I.; Rothuizen, H. E.; Stutz, R.; Widmer, R.; Binnig, G. K. The "Millipede" - More than one thousand tips for future AFM data storage. *IBM J. Res. Dev.* **2000**, *44*, 323-340.

42. Kim, F.; Sohn, K.; Wu, J.; Huang, J. Chemical synthesis of gold nanowires in acidic solutions. *J. Am. Chem. Soc.* **2008**, *130*, 14442-14443.

43. Jana, N. R.; Gearheart, L.; Murphy, C. J. Wet chemical synthesis of high aspect ratio cylindrical gold nanorods. *J Phys. Chem. B* **2001**, *105*, 4065-4067.

44. Horcas, I.; Fernandez, R.; Gomez-Rodriguez, J. M.; Colchero, J.; Gomez-Herrero, J.; Baro, A. M. WSxM: A software for scanning probe microscopy and a tool for nanotechnology. *Rev. Sci. Instrum.* **2007**, *78*, 013705.

45. Gimeno, A.; Ares, P.; Horcas, I.; Gil, A.; Gomez-Rodriguez, J. M.; Colchero, J.; Gomez-Herrero, J. 'Flatten plus': a recent implementation in WSxM for biological research. *Bioinformatics* **2015**, *31*, 2918-2920.

46. Liu, H.-Z.; Wu, S.; Zhang, J.-M.; Bai, H.-T.; Jin, F.; Pang, H.; Hu, X.-D. Strategies for the AFM-based manipulation of silver nanowires on a flat surface. *Nanotechnology* **2017**, *28*, 365301.

# Supporting Information

## Table of Contents

## SI1. Nanowires manipulation

Figure S1-1 depicts the steps comprising the Au NWs manipulation with an AFM probe. After imaging the sample in amplitude modulation mode (AM-AFM), we bring the tip down to the substrate surface applying a normal/lateral load of about 60 to 300 nN (Fig. S1-1a) and we move it along a predefined trajectory (Fig. S1-1b), thus manipulating the Au NW of our choice (Fig. S1-1c). After this, we lift the tip back to AM-AFM mode and we image the results of the manipulation (Fig. S1-1d).

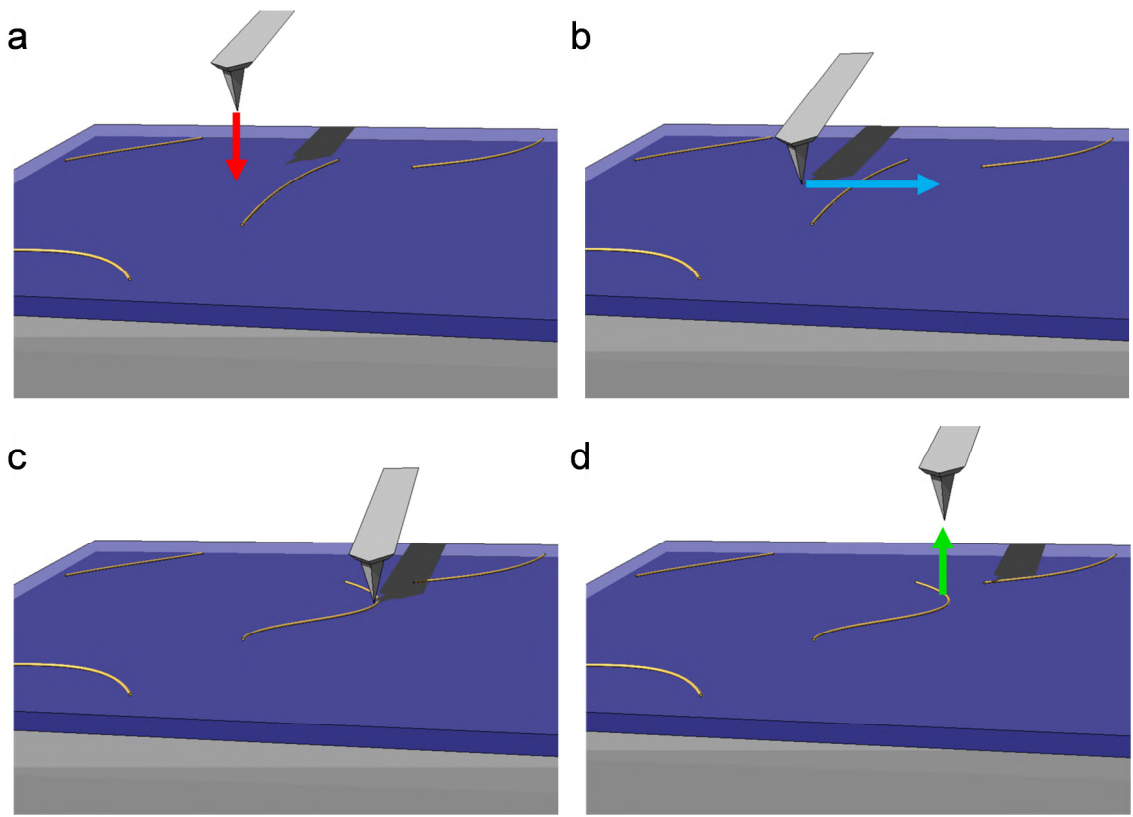

**Figure S1-1. Au NWs manipulation schematics. a**, After imaging Au NWs in AM-AFM mode, we bring the tip into contact. **b**, We move the tip along a predefined trajectory to manipulate the Au NW. **c**, The selected Au NW is moved. **d**, We bring the tip back to AM-AFM mode and a new image is taken.

Several trajectories can be performed in a single nanomanipulation step to translate/rotate the NWs. Figure S1-2 shows an example of these nanomanipulations and the corresponding script on the control software,[1] where a remarkably simple procedure is employed. Additionally, Fig. S1-3 depicts the temporal traces of the normal and lateral forces while two NWs are displaced by means of 5 movements of the tip in contact with the substrate.

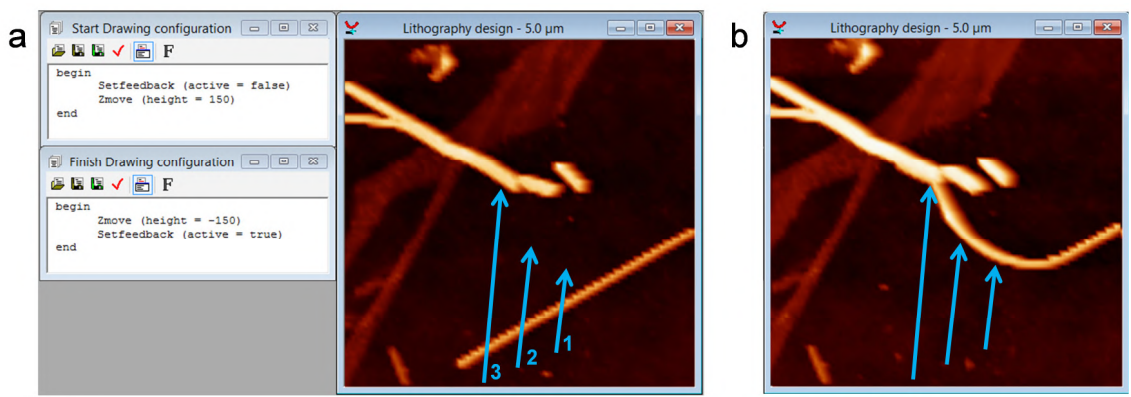

**Figure S1-2. Example of Au NWs assembly. a**, Script (left) and AFM topographic image of some Au NWs before manipulation (right). **b**, AFM topographic image of the same area as in **a** after the nanomanipulation. Blue arrows represent the tip trajectories in contact mode used to bring the NW in the bottom into contact with the one in the middle of the image. Numbers beside the arrows indicate the order of the manipulations.

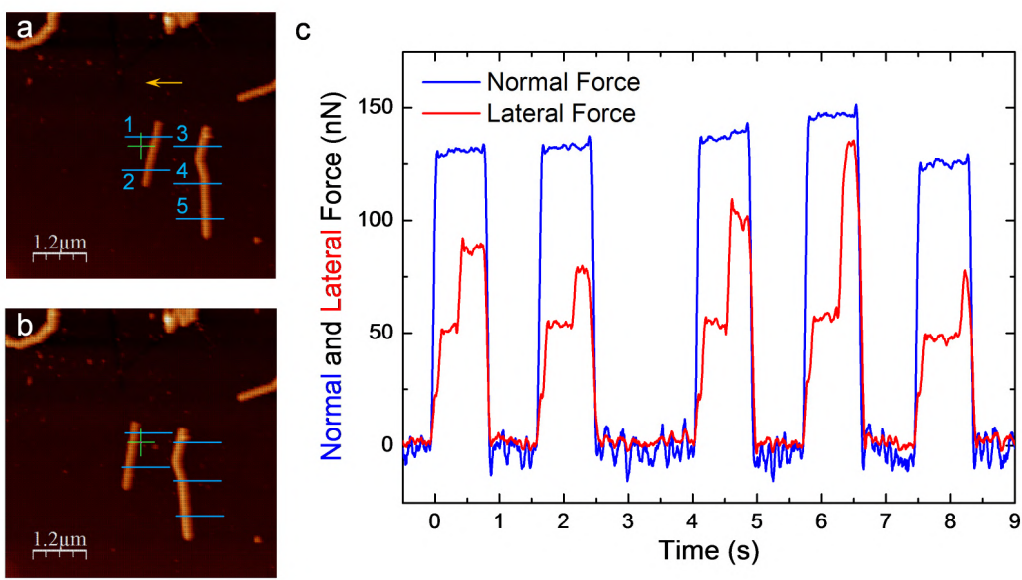

**Figure S1-3. Example of Au NWs manipulation and monitorization of the normal and lateral forces . a**, **b,** AFM topographies of two NWs before and after manipulation. The trajectories that the tip follows in contact with the substrate are depicted with blue lines. Numbers beside each line indicate the order of the manipulations. The yellow arrow shows the direction of the trajectories. **c**, Evolution of the normal (blue line) and lateral (red line) forces during the manipulation. Both forces are obtained in nN following usual AFM calibration methods. The manipulation was performed on silicon dioxide at air ambient conditions.

The value of the normal/lateral force needed to move a NW depends on a number of factors, being the most important one the interaction of the wire with the substrate. Moving a NW for the first time usually requires a higher normal force (200-400 nN) than that needed for subsequent movements (which can be performed with only 60 nN). The orientation of the NW with respect to the direction of movement also plays an important role: many NWs can be moved with lower forces in certain directions.
The NWs which are adjacent to gold islands (which usually have triangular or hexagonal shapes) are hardly movable. Regardless previous factors, a few NWs are kind of "anchored" to the substrate and cannot be moved or they require much higher loads.

Following Lu *et al.* [2] to obtain an optimum electrical contact between two NWs, one of them is pushed against the other and they will cold weld just by mechanical contact. In Lu *et al.* work, they cold weld Au NWs with diameters between 3 and 10 nm. The authors used different relative orientations between the NWs: (i) "head-to-head" (parallel NWs "touching" each other at the end), (ii) "side-to-side" (parallel NWs laterally "touching" each other) and (iii) "head-to-side" (perpendicular NWs, the end of one of them "touches" one side of the other). In that work any configuration was successful, but in our case, we manipulate much higher diameters (44 - 84 nm) and the "head-to-side" geometry was the one providing better electrical joining. Furthermore, it is the one most easily accessible from a manipulation point of view. Anyhow, we also obtained some good electrical junctions with the other two configurations.

## SI2. Electron beam lithography (EBL) contacts on few layer antimonene

Figure S2-1 portrays an optical microscope image of a nanocircuit fabricated by EBL. The inset is an AFM topography showing a zoom in image of the region where four gold electrodes connect a few layer antimonene flake. We prepared the circuit following standard EBL procedures. We revised the connecting paths thoroughly, but we were not able to measure electrical current above the sensitivity of equipment. Similar experiments were carried out three times with different samples and same results.

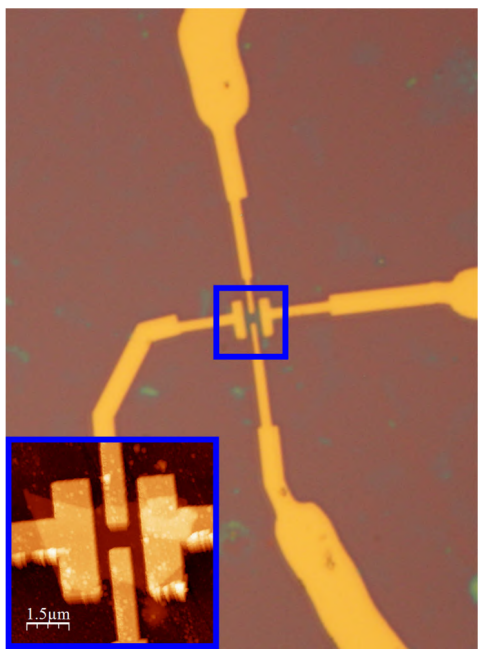

**Figure S2-1. Electron Beam Lithography contacts on few layer antimonene.** Optical image of a nanocircuit fabricated by EBL. The inset is an AFM topography showing a magnified image of the central region encircled in blue, where the target flake can be readily seen.

## SI3. Nanoelectrode reconfiguration

In addition to the example described in Fig. 2 and 3 in the manuscript where we used the conductive AFM configuration, we show here another experiment, this time using few layer graphene flakes. Figure S3-1a shows an AFM topography where several Au NWs protruding from a gold microelectrode (right) are clearly seen. Figure S3-1b shows a gold NW path connecting the lower graphene flake and Fig. S3-1c the corresponding *IV* characteristics at the red and blue crosses respectively. After reconfiguring the circuit, we connected the upper flake (Fig. S3-1d) and we measured the resistance *vs.* length dependence with a metallized AFM tip following the white crosses drawn in Fig. S3-1d. Figure S3-1e shows the result of the measurements. From this plot we infer a contact resistance of ~9 k$\Omega$ and taking into account the dimensions and geometry of the few layer graphene flake we obtain a sheet resistivity of 670 ± 60 $\Omega$/sq.

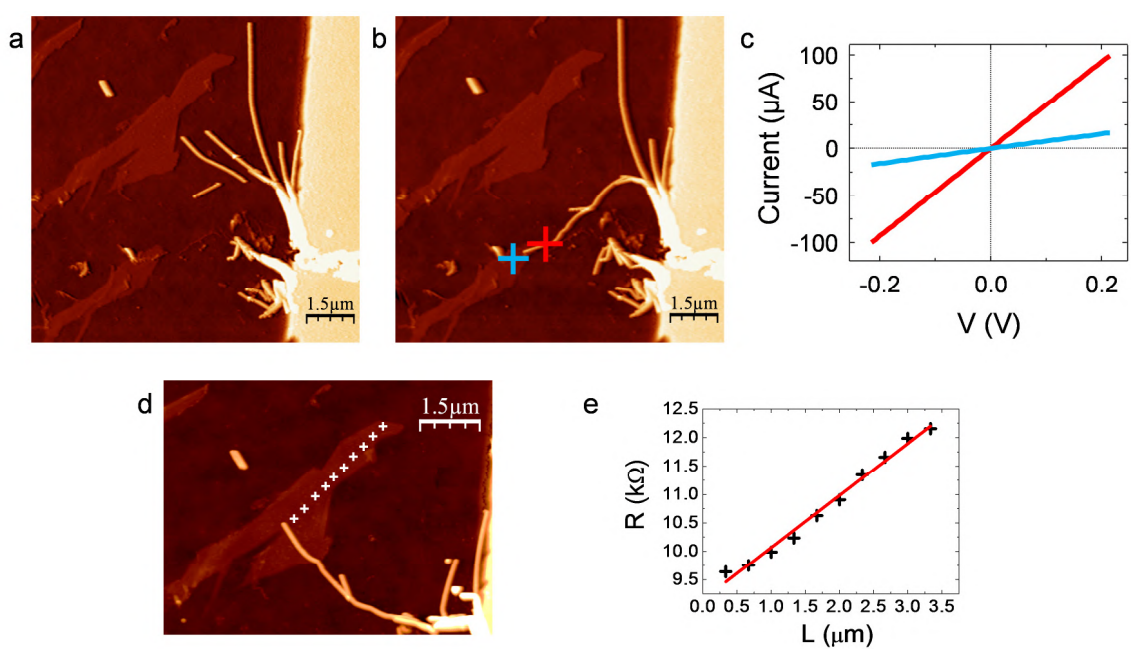

**Figure S3-1. Circuit reconfiguration in graphene. a**, AFM topography showing two few layer graphene flakes (left), gold nanowires and a gold microelectrode (right). **b**, Connection with the lower flake. **c**, Current *vs.* voltage characteristics in the red and blue crosses drawn in **b**. **d**, Connection to the upper graphene flake. **e**, Resistance *vs.* length dependence taken along the trajectory marked by the white crosses in **d**.

Figure S3-2 shows several of the different nanoelectrode configurations used to measure the resistance of the multiwalled carbon nanotube (MWCNT) referred in the main text as a function of the distance between the two Au NW electrodes contacting it.

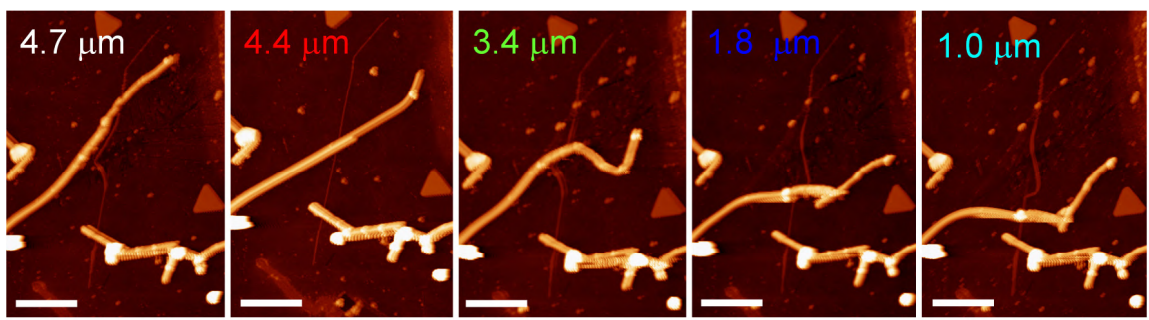

**Figure S3-2. Nanoelectrode reconfiguration along the MWCNT mentioned in the main text.** Locations where we acquired *IV* curves along the MWCNT experiment. The distance between the Au NW electrodes along the MWCNT is shown in the top left corner of each AFM topography image. The color corresponds to the color of the dots in Fig. 4e in the main text. Scale bar length: 2 microns.

Figure S3-3 shows the *IV* and the current *vs.* gate voltage curves acquired at the carbon nanotube device of the main text. These measurements were performed when the distance between the gold nanowires connecting the nanotube was 4.7 µm. Part a presents an almost linear response of the drain-source current *vs.* bias voltage, while part b shows a strong hysteresis with backgate voltage. This hysteresis is quite common in this kind of devices and it has been attributed to charge trapping by water molecules around the nanotubes, including $SiO_2$ surface-bound water close to the nanotubes.[3]

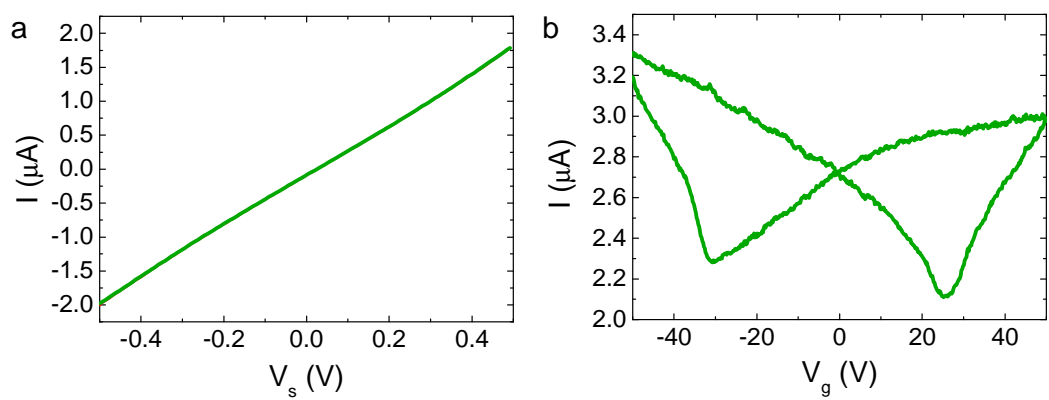

**Figure S3-3. Electrical characterization of a MWCNT electrically contacted via SPANC.** Performed in air at ambient conditions. **a**, Drain-source current as a function of the bias voltage (at a gate voltage of 0 V) for the nanocircuit of figure 4a. **b**, Current *vs.* gate voltage dependence in the same device with a bias voltage of 0.5 V.

## SI4. Gold nanoparticles device fabrication using SPANC

We deposited Au NPs by drop-casting following a similar approach as to deposit Au NWs. On a substrate where Au microelectrodes were evaporated, we first deposited Au NWs (Fig. S4-1a). Then, we deposited Au NPs (Fig. S4-1b). A direct comparison of the images before and after the Au NPs deposition allows us to clearly identify the location of the nanoparticles.

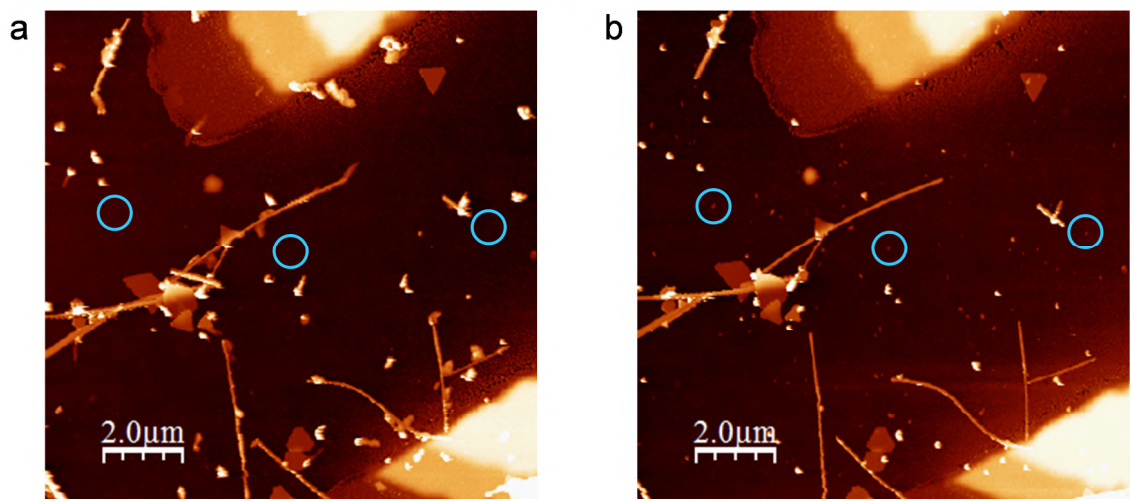

**Figure S4-1. Gold nanoparticles deposition. a**, AFM Topographic image of an area with two gold microelectrodes and Au NWs. **b**, Same area as in **a**, but after Au NPs deposition. We have encircled 3 NPs to guide the eye. In **a** the circles are empty while in **b** they contain nanoparticles.

Figure S4-2 shows and intermediate stage of the nanomanipulation where the capture of the bottom NP of the device can be readily seen.

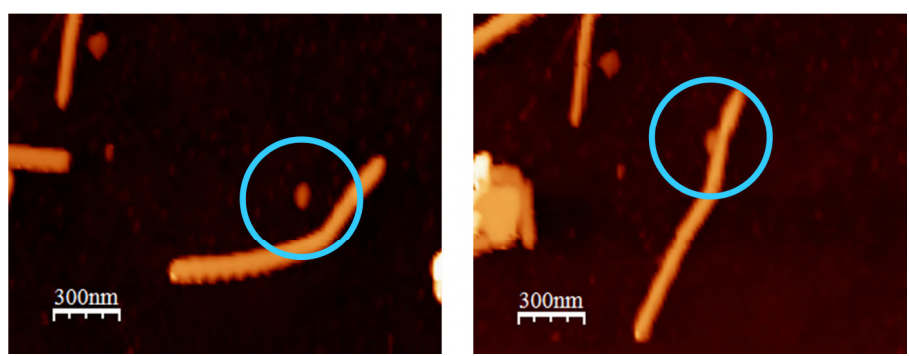

**Figure S4-2. Capture of a gold nanoparticle for device assembly.** Detail of the Au NWs manipulation and capture of an Au NP for its electrical contact. We have encircled the nanoparticle to guide the eye.

# SI5. Measurement of the 1,4-benzenedithiol molecule conductivity using SPANC

Once the two nanowires were as close as possible avoiding cold welding (confirmed by the absence of current between electrodes), we used a sharp tip to check the gap width finding it within the tip resolution. After all the 1,4-benzenedithiol (BDT) measurements, we used a multiwalled carbon nanotube (MWCNT) sample to check the final radius of the same sharp tip. Figure S5-1a shows a MWCNT with a height of h ~15 nm and width W ~38 nm (see the corresponding profile in Fig. S5-1b). From these two magnitudes, we obtained a tip radius of ~10 nm calculated as $R_{tip} = W^2/8h$ (following Ref. [4]).

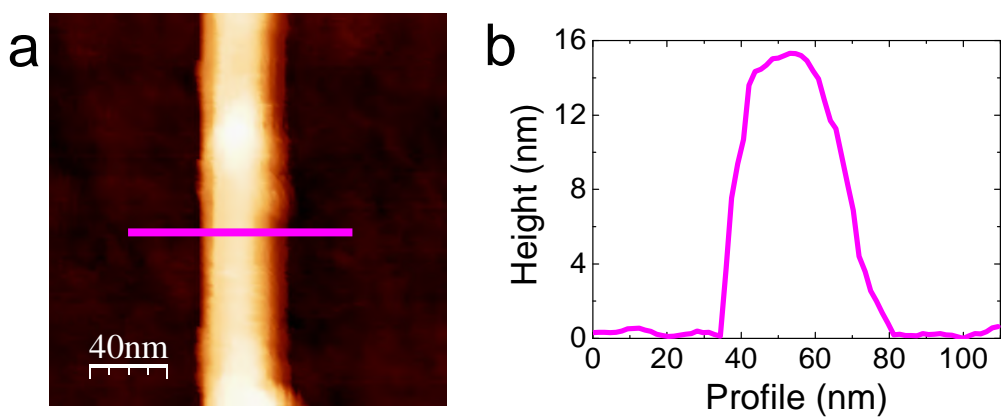

**Figure S5-1. Tip radius calibration. a**, AFM topographic image of a multiwalled carbon nanotube used to obtain the tip radius value. **b**, Height profile along the line in **a**.

After BDT deposition on the device, we confirmed the absence of cold welding by measuring current *vs.* voltage characteristics showing no measurable current for bias voltages between 0 and 4.5 V. We performed *IV*s at higher bias voltages until we observed the onset of current for a bias voltage of 5 V (see inset in Fig. S5-2). Figure S5-2 shows *IV*s acquired at lower voltages, roughly obtaining the same conductance. For bias voltages higher than 0.8 V, peaks in the current appeared, a common feature in molecular electronics transport measurements.[5]

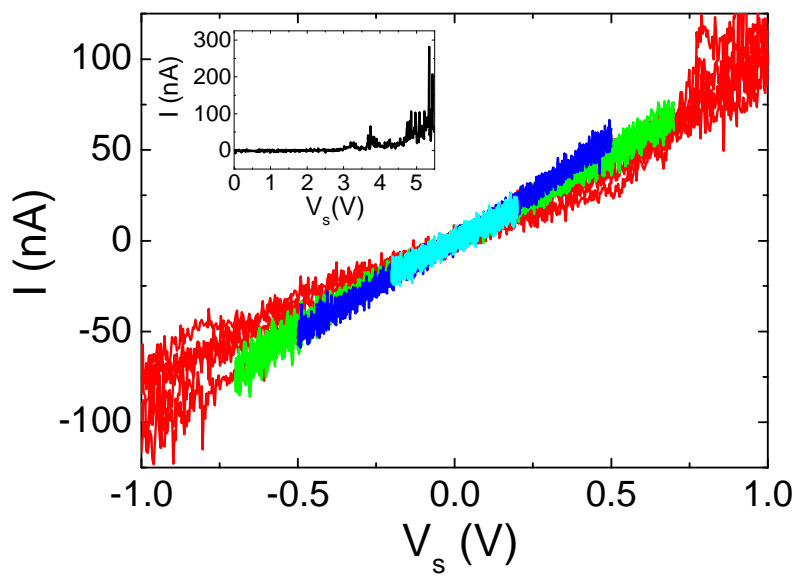

**Figure S5-2. Current *vs.* voltage characteristics carried out after 1,4-benzenedithiol deposition**. The inset shows an asymmetric IV where the onset of the current is observed when reaching a bias voltage of ~5 V.

Once a good electrical NW-BDT-NW contact was achieved, we carried out additional *IV*s increasing the bias voltage range, observing a final drop to almost zero resistance that we interpret as a cold welding event. Figure S5-3 shows a collection of representative *IV*s with increasing bias voltage.

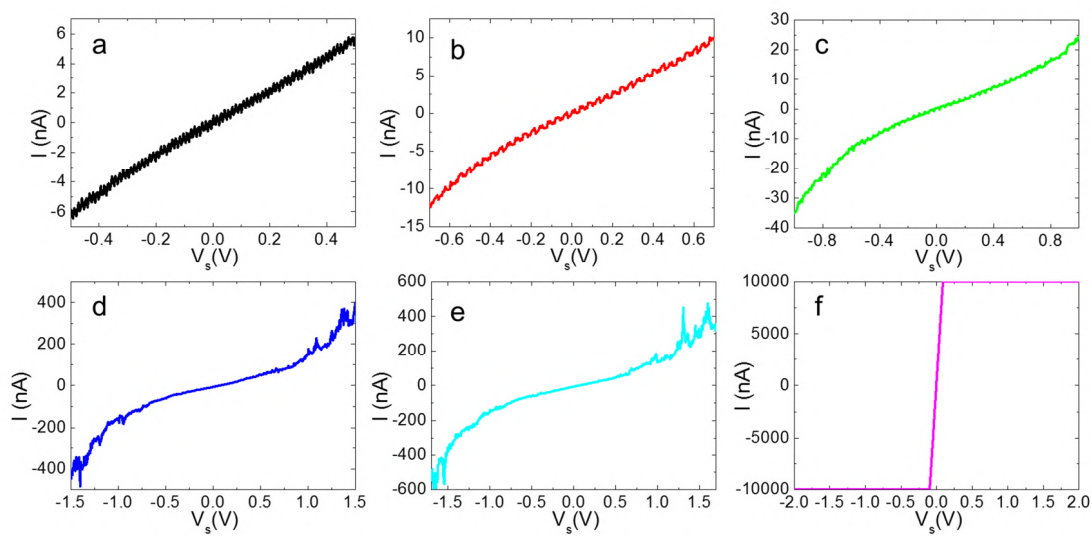

**Figure S5-3. *IV* characteristics of 1,4-benzenedithiol molecules contacted by SPANC with increasing bias voltage. a**, ± 0.5 V. **b**, ± 0.7 V. **c**, ± 1.0 V. **d**, ± 1.5 V. **e**, ± 1.7 V. **f**, ± 2.0 V where the cold welding finally takes place.

Figure S5-4 shows a transmission electron microscope (TEM) image of two adjacent nanowires. According to this image, the minimum gap width between NWs is about 1.6 nm. We attribute this distance to the presence of CTAB surfactant that prevents NWs coalescence. This further confirms our assumption for the ~1 nm gap in Fig. 6 in the main text.

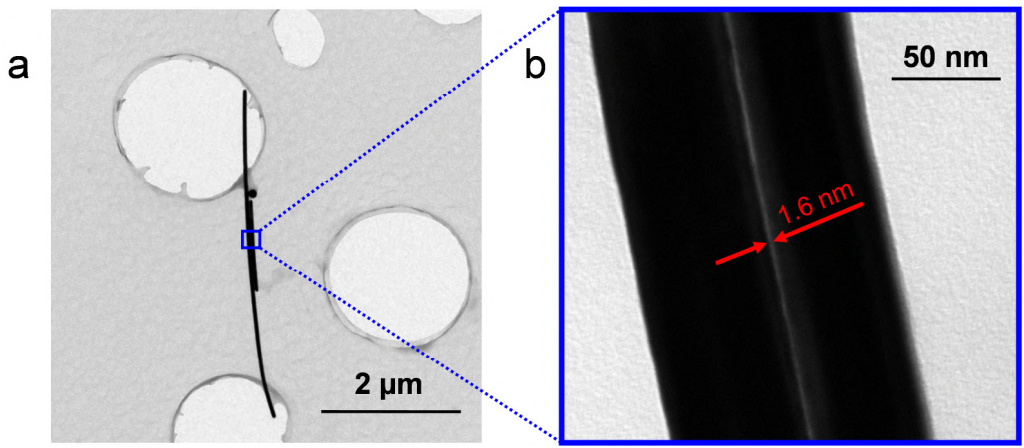

**Figure S5-4. TEM images of Au NWs. a**, Large area image showing two adjacent NWs. **b**, Zoom image showing the region enclosed in **a**. A gap of about 1.6 nm can be readily seen.

A representative conductance trace measured once a NW-BDT-NW junction is achieved is displayed in Fig. S5-5a. This trace has two parts. The first one, from 20 s to around 37 s, reveals that the conductance is switching between two values. The second part (for time > 37s) also displays two levels for the conductance; in this case, the lower one matches the upper one of the first part of the plot. The right panel in Fig. S5-5a shows the corresponding conductance histogram. The three values of the conductance are evidenced by the three visible peaks. The higher number of counts of the central peak suggests enhanced stability around that conductance value.

The obtained switching behavior between two values (with a superimposed background noise in our data) is characteristic of the so-called random telegraph signal (RTS) or bi-stable noise.[6] This telegraphic noise occurs once a molecular junction is formed,[5] and it is attributed to spontaneous binding and unbinding of molecules to the electrodes (the NWs in this experiment). The power spectral density (PSD) of the trace is depicted in Fig. S5-5b disclosing the expected shape for a RTS: a plateau for low frequencies followed by a $1/f^2$ decay for high frequencies. This plot further supports our results.

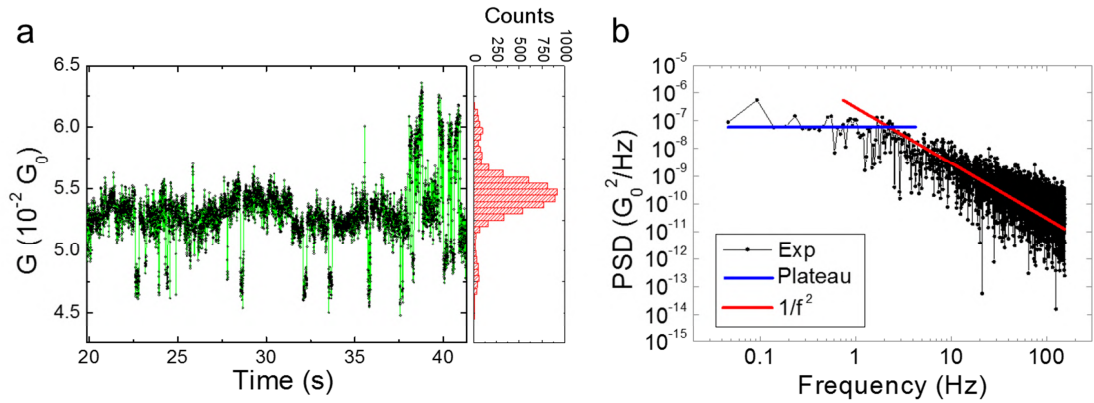

**Figure S5-5. Temporal and spectral analysis of a conductance trace of NW-BDT-NW junction. a**, Conductance temporal trace (data plotted with black points connected by a green line). The right panel is the corresponding histogram, showing a wide high peak between two lower peaks, which correspond to the central most probable value of the conductance and the two other levels, respectively. **b**, Power spectral density (PSD) of the trace shown in **a** together with the corresponding fitting to a plateau and $1/f^2$, characteristic of RTS spectra.

## SI6. Manipulation and welding of Au NWs on mica

We have fabricated devices with electrodes and nanowires on a mica substrate. Mica is an interlayer compound exhibiting a flat surface with a quite different chemical nature from that of silicon dioxide (the typical substrate used in our experiments). We have performed manipulation and cold welding of nanowires in air ambient conditions (40 % RH, T ~ 26°C) finding no significant differences respect to what we observed in silicon dioxide.

Figure S6-1 shows a sequence of AFM topography images of the manipulation of the NWs from their initial distribution until the final configuration that consists in a path of three NWs.

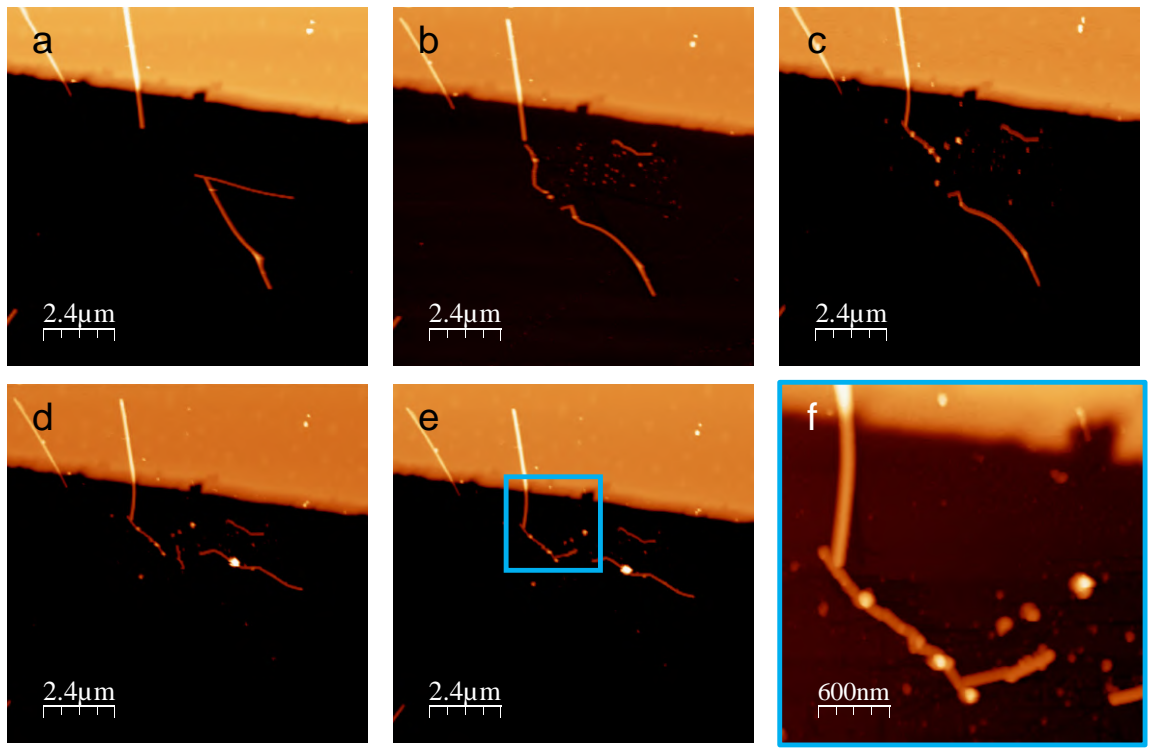

**Figure S6-1. Manipulation of Au NWs on mica at air ambient conditions. a-e**, Sequence of AFM topography images of the manipulation. **f**, Zoom in on the path composed of three NWs, marked with a blue square in **e**.

To demonstrate the successful cold welding between nanowires under these conditions, we acquired an *IV* curve on the last NW of the path using a doped diamond AFM tip to close the circuit. We connected a protection resistor of 4.7 kΩ in series, this resistance together with the tip-Au NW contact resistance accounts for the value of the resistance of this *IV* curve. Obtaining the *IV* curve at the end of the path (green point in Fig. S6-2a and green curve in Fig. S6-2b) proves the cold welding between the NWs.

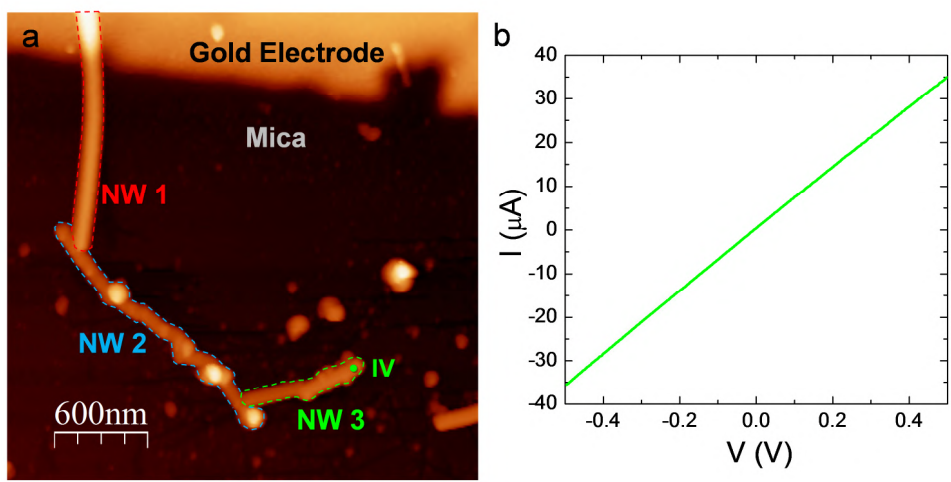

**Figure S6-2. Checking the cold welding between NWs manipulated on mica at air-ambient conditions. a**, AFM topography of the nanoelectrode formed by three nanowires. **b**, *IV* curve acquired at the end of the path of NWs. A protection resistor of 4.7 kΩ was connected in series.

## SI7. Manipulation and welding of Au NWs at different humidities

By modifying our experimental set up we were able to measure from medium vacuum ($10^{-2}$ mbar) up high humidity (80 ± 5 % RH, T = 26.5 ± 0.5°C). For no particular reason, we used mica as substrate. Again, we were able to manipulate the nanowires to fabricate electrical circuits without any appreciable difference respect to the experiments previously carried out at ambient humidity. Regarding the medium humidity range, we have performed experiments in Madrid from spring to winter. This covers a humidity range from 30 % up to 55 % RH. In summary, these experiments characterize the fabrication of circuits with nanowires by AFM from very low to very high humidity values.

**Medium vacuum.** The main results of the manipulation carried out in medium vacuum are summarized in Figs. S7-1 and S7-2, which show respectively the manipulation of the nanowires and the verification of the cold welding between them. When manipulating the nanowires the only difference with respect to the manipulation at air ambient conditions was the AFM mode used to image the nanowires: instead of using Amplitude Modulation (AM) dynamic mode, we imaged the sample in Drive Amplitude Modulation (DAM) dynamic mode.[7]

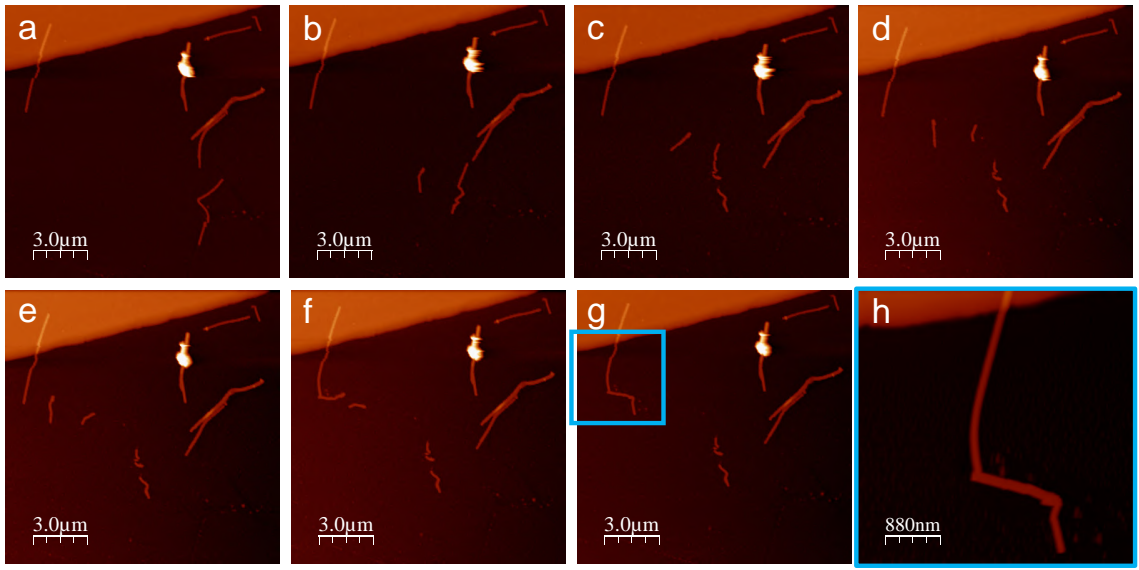

**Figure S7-1. Manipulation of Au NWs on mica at medium vacuum. a-g**, Sequence of AFM topography images of the manipulation. **h**, Zoom in on the path composed of three NWs, marked with a blue square in **g**.

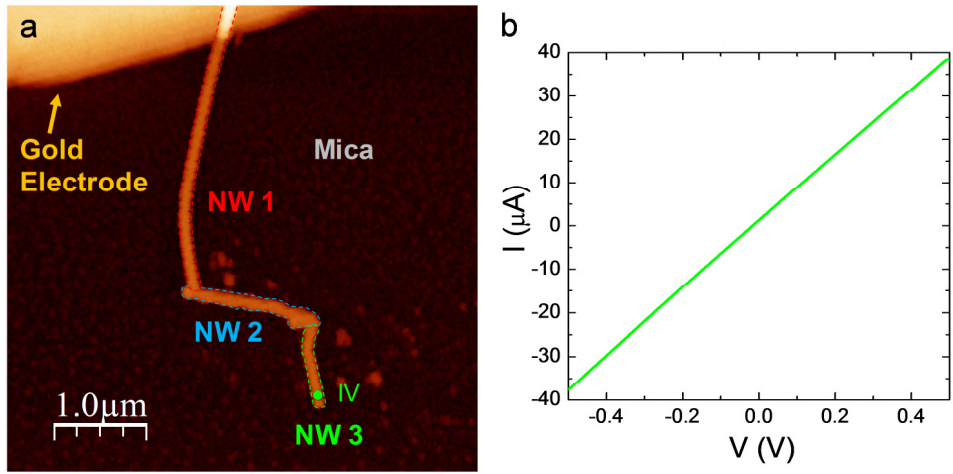

**Figure S7-2. Checking the cold welding between NWs manipulated on mica at medium vacuum. a**, AFM topography of the nanoelectrode formed by three nanowires. **b**, *IV* curve acquired at the end of the path of NWs. A protection resistor of 4.7 kΩ was connected in series.

Figure S7-3 shows the evolution of the normal and lateral forces while the AFM tip is manipulating a NW at medium vacuum (~ 0 % RH and T ~ 25°C).

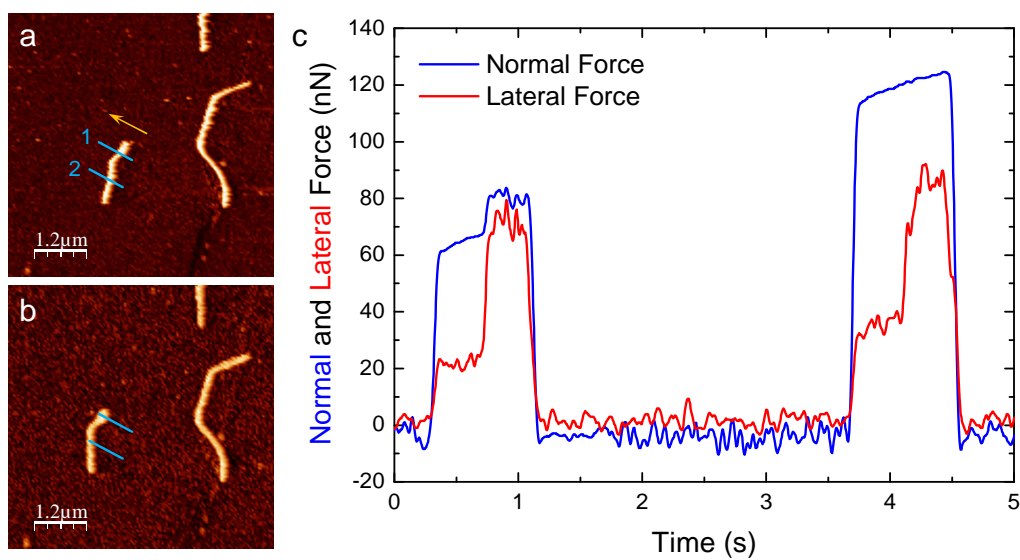

**Figure S7-3. Au NWs manipulation and monitorization of the normal and lateral forces on mica at medium vacuum. a,b,** AFM topographies of NWs before and after the manipulation. The trajectories that the tip follows in contact with the substrate are depicted with blue lines. Numbers beside each line indicate the order of the manipulations. The yellow arrow shows the direction of the trajectories. **c,** Evolution of the normal (blue line) and lateral (red line) forces during the manipulation. Both forces are obtained in nN following usual AFM calibration methods.

**High Humidity.** Figure S7-4 shows the manipulation of NWs performed at high humidity (80 ± 5 % RH, T = 26.5 ± 0.5°C) while Fig. S7-5 reveals the electrical continuity of the path formed by the manipulation.

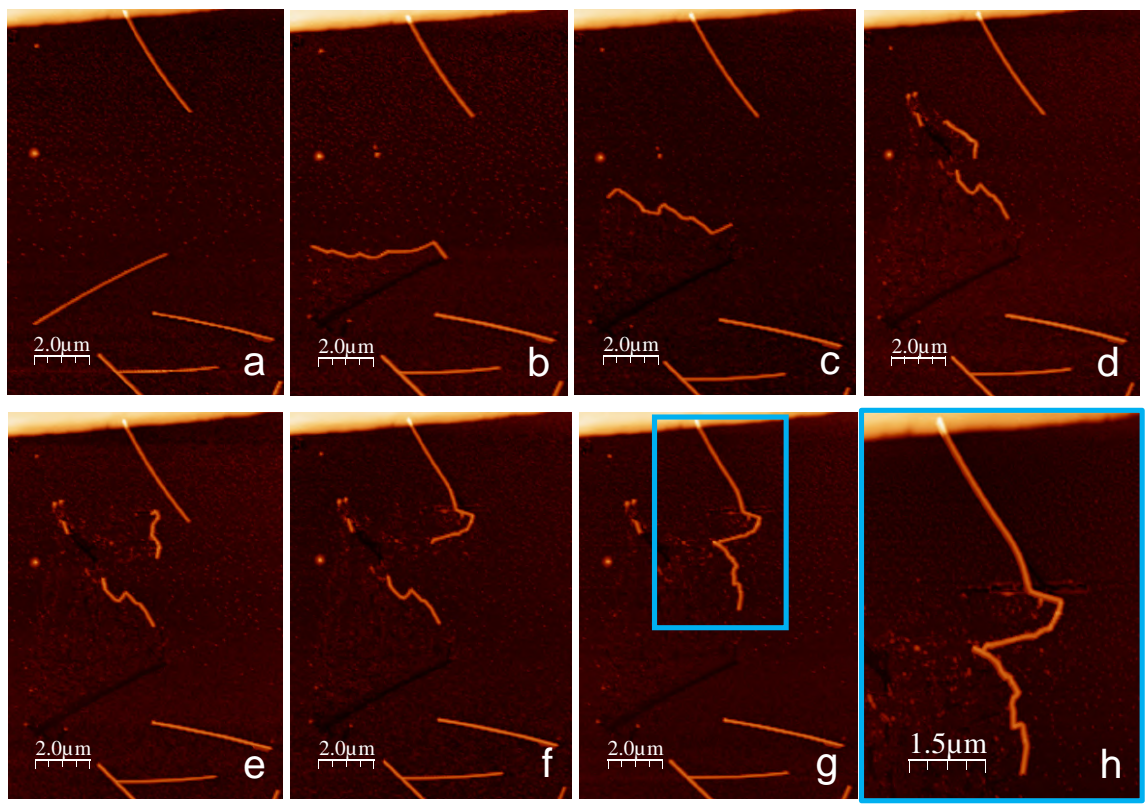

**Figure S7-4. Manipulation of Au NWs on mica at high humidity. a-g**, Sequence of AFM topography images of the manipulations. **h**, Zoom in on the path composed of three NWs, marked with a blue square in **g**.

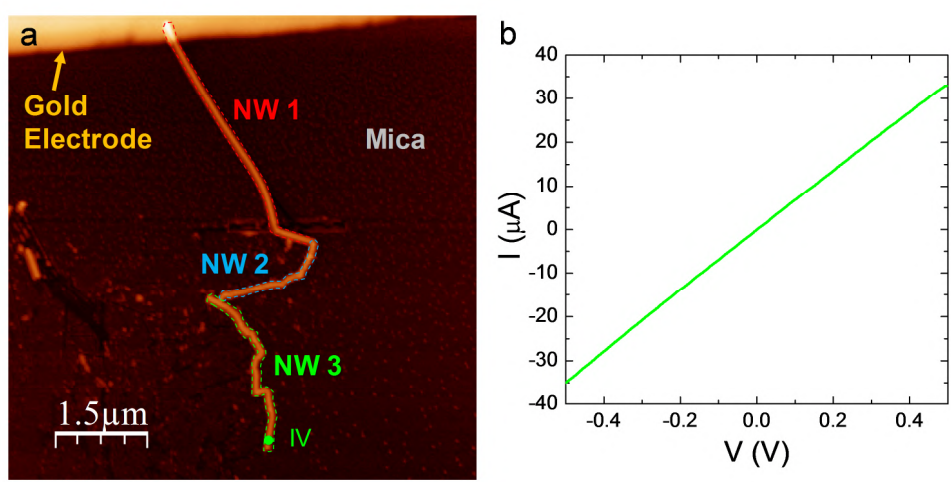

**Figure S7-5. Checking the cold welding between NWs manipulated on mica at high humidity. a**, AFM topography of the nanoelectrode formed by three nanowires. **b**, *IV* curve acquired at the end of the path of NWs. A protection resistor of 4.7 kΩ was connected in series.

## SI8. Preparation of gold nanowires

**Chemicals**. Gold(III) chloride hydrate (99.999 %) was purchased from Aldrich, hexadecyltrimethylammonium bromide (CTAB, 99 %) from Sigma, sodium borohydride, tri-sodium citrate 2-hidrate (99 %) and L(+)-ascorbic acid (90 %) were purchased from Panreac. All synthetic reagents were used as received. Milli-Q water was used in all synthesis. All the glassware must be washed with aqua regia and rinsed thoroughly with Milli-Q water.

**Preparation of seed gold nanoparticles.** The colloidal solution was prepared according to standard literature procedure[8]. To a 10 mL aqueous solution containing $HAuCl_4$ (0.25 mM) and trisodium citrate (0.25 mM), prepared in a 30 mL glass vial, 0.3 mL of ice cold 0.1 M $NaBH_4$ solution were added under stirring. The solution turned orange immediately after addition. These nanoparticles can be used as seeds after 5 h preparation and can be stored at 4 °C.

**Preparation of gold nanowires.** A three step seeding method was used for nanowire preparation. 30 mL of the growth solution were prepared by adding 15 mL of chloroauric acid solution (0.5 mM) to 15 mL of CTAB (0.2 M). Then, two 20 mL glass vials (labeled *A* and *B*) and a 50 mL PP tube labeled *C* were prepared. Vials *A* and *B* containing 2.25 mL of growth solution were mixed with 12.5 µL of 0.1 M ascorbic acid solution. Tube *C*, containing 23 mL of growth solution, was mixed with 130 µL of 0.1 M ascorbic acid solution and 100 µL of nitric acid. Next, 200 µL of 4 nm seed solution were mixed with sample *A* and stirred, the color of *A* turned pink. After 5 s, 200 µL of *A* were added to sample B followed by thorough mixing for 5 seconds. The color of *B* turned light pink. After 5 s, 200 µL of *B* were added to *C* and stirred during 10 s. The nanowires were allowed to grow during 48 h without stirring at 28.5 °C, this temperature is essential in order to avoid CTAB precipitation problems.

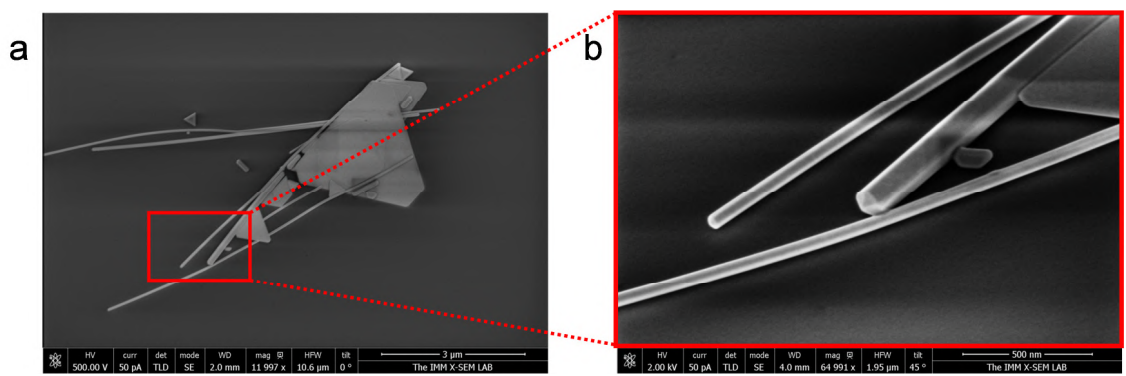

**Figure S8-1. Scanning electron microscope (SEM) representative images of the single crystal Au NWs obtained following the above described procedure. a**, Image showing several Au NWs and Au islands. **b**, Zoom in of the region enclosed by the red rectangle in **a**, where the pentagonal cross-section of the NWs can be seen.

## SI9. Stencil masks for thermal evaporation

We ordered specially designed masks made of Nickel (Gilder Grids Ltd.). Figure S9-1 shows the layout of the masks. Each mask contains 8 motifs that were designed to allow the fabrication of four and six microelectrode setups with gaps ranging between 50 and 5 µm.

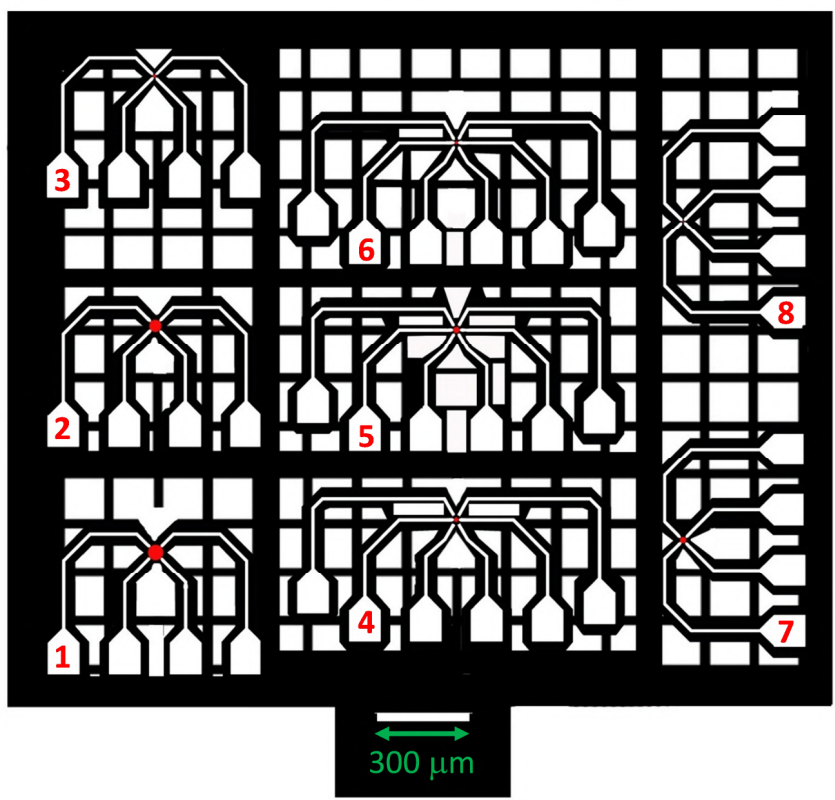

**Figure S9-1. Stencil masks layout.** Nickel mask design with the following gap sizes in each motif in µm: 1) 50, 2) 40, 3) 20, 4) 15, 5) 10, 6) 6, 7) 10 and 8) 5.

We considered two different types of samples: a) inhomogeneous, the nano-objects of interest are located by inspection, usually with an optical microscope. In this case, the sample substrate is glued to a strong Cobalt-Samarium magnet and the gap of one mask motif is located on top of the object with a micromanipulator. As the mask is made out of Nickel, it remains firmly attached to the substrate thanks to the magnetic force of the magnet. Then, the magnet with the sample and the attached mask are inserted on a thermal evaporator. b) Homogeneous sample. The nano-objects cover the sample uniformly. In this case, there is no need to locate the mask in any particular region of the substrate.

# References

1. Horcas, I.; Fernandez, R.; Gomez-Rodriguez, J. M.; Colchero, J.; Gomez-Herrero, J.; Baro, A. M.WSXM: A software for scanning probe microscopy and a tool for nanotechnology. *Rev. Sci. Instrum.* **2007,** 78, 013705.
2. Lu, Y.; Huang, J. Y.; Wang, C.; Sun, S. H.; Lou, J.Cold welding of ultrathin gold nanowires. *Nat. Nanotechnol.* **2010,** 5, 218-224.
3. Kim, W.; Javey, A.; Vermesh, O.; Wang, O.; Li, Y. M.; Dai, H. J.Hysteresis caused by water molecules in carbon nanotube field-effect transistors. *Nano Letters* **2003,** 3, 193-198.
4. Markiewicz, P.; Goh, M. C. Atomic-force microscopy probe tip visualization and improvement of images using a simple deconvolution procedure. *Langmuir* **1994,** 10, 5-7.
5. Brunner, J.; Teresa Gonzalez, M.; Schoenenberger, C.; Calame, M.Random telegraph signals in molecular junctions. *J. Phys.: Condens. Matter* **2014,** 26, 474202.
6. Yuzhelevski, Y.; Yuzhelevski, M.; Jung, G.Random telegraph noise analysis in time domain. *Review of Scientific Instruments* **2000,** 71, 1681-1688.
7. Jaafar, M.; Martinez-Martin, D.; Cuenca, M.; Melcher, J.; Raman, A.; Gomez-Herrero, J.Drive-amplitude-modulation atomic force microscopy: From vacuum to liquids. *Beilstein Journal of Nanotechnology* **2012,** 3, 336-344.
8. Jana, N. R.; Gearheart, L.; Murphy, C. J.Wet chemical synthesis of high aspect ratio cylindrical gold nanorods. *Journal of Physical Chemistry B* **2001,** 105, 4065-4067.